\documentclass[twocolumn]{aastex631}
\usepackage{amsmath}
\usepackage{multirow}
\usepackage{bigdelim}
\graphicspath{{figures/}}
\usepackage{listings}
\usepackage{xcolor}

\definecolor{codegreen}{rgb}{0.2902, 0.5412, 0.3059}
\definecolor{codeblue}{rgb}{0.3059, 0.5059, 0.8314}
\definecolor{codepurple}{rgb}{0.58, 0, 0.82}
\definecolor{backcolour}{rgb}{0.98,0.98,0.98}
\lstdefinestyle{codestyle}{
    backgroundcolor=\color{backcolour},
    commentstyle=\color{codegreen},
    keywordstyle=\color{codeblue},
    showstringspaces=false,
    stringstyle=\color{codepurple},
    basicstyle=\ttfamily\footnotesize,
}
\newcommand{\norm}{\mathcal{N}}

\newcommand{\probb}{p(\mathcal{B}|y)}
\newcommand{\nss}{n_\sigma^*}

\begin{document}
\title{Bayesian Model Comparison and Significance: Widespread Errors and how to Correct Them}
\author[0000-0002-5113-8558]{Daniel P. Thorngren}
\affiliation{William H. Miller III Department of Physics \& Astronomy, Johns Hopkins University, 3400 N Charles St, Baltimore, MD 21218, USA}
\author[0000-0001-6050-7645]{David K. Sing}
\affiliation{William H. Miller III Department of Physics \& Astronomy, Johns Hopkins University, 3400 N Charles St, Baltimore, MD 21218, USA}
\affiliation{Department of Earth and Planetary Science, Johns Hopkins University, 3400 N. Charles Street, Baltimore, MD 21218, USA}
\author[0000-0003-1622-1302]{Sagnick Mukherjee}
\affiliation{Department of Astronomy and Astrophysics, University of California, Santa Cruz, Santa Cruz, CA 95064, USA}
\affiliation{William H. Miller III Department of Physics \& Astronomy, Johns Hopkins University, 3400 N Charles St, Baltimore, MD 21218, USA}

\begin{abstract}
Bayes factors have become a popular tool in exoplanet spectroscopy for testing atmosphere models against one another.  We show that the commonly used method for converting these values into significance ``sigmas" is invalid.  The formula is neither justified nor recommended by its original paper, and overestimates the confidence of results.  We use simple examples to demonstrate the invalidity and prior sensitivity of this approach.  We review the standard Bayesian interpretation of the Bayes factor as an odds ratio and recommend its use in conjunction with the Akaike Information Criterion (AIC) or Bayesian Predictive Information Criterion Simplified (BPICS) in future analyses (Python implementations are included) .  As a concrete example, we refit the WASP-39 b NIRSpec transmission spectrum to test for the presence of SO$_2$.  The prevalent, incorrect significance calculation gives $3.67\sigma$ whereas the standard Bayesian interpretation yields a null model probability $\probb={0.0044}$.  Surveying the exoplanet atmosphere literature, we find widespread use of the erroneous formula.  In order to avoid overstating observational results and estimating observation times too low, the community should return to the standard Bayesian interpretation.
\end{abstract}

\section{Introduction}
The use of Bayesian statistics has become common practice in the analysis of exoplanetary spectra \citep[e.g.][]{Madhusudhan2011a, Benneke2012, Gibson2012, Kreidberg2015a, Spake2021, Ahrer2023}.  These often make use of Markov Chain Monte-Carlo (MCMC) to sample the posterior distribution of various planetary parameters and atmospheric abundances given the observed data, using tools such as EMCEE \citep{Foreman-Mackey2013} or Dynesty \citep{Speagle2020, Koposov2024}.  In planetary science, these have been termed ``atmospheric retrievals" and have played a key role in major exoplanet observational results in the field \citep[e.g.][]{Fraine2014, Kreidberg2014, Benneke2019, Piaulet-Ghorayeb2024}.  While there are a variety of Bayesian approaches for considering this data, this letter is mainly concerned with the use of Bayes factors in model comparison \citep[see][]{Jeffreys1935,Kass1995}.

The Bayes factor approach is based on the normalizing denominator term $p(y)$ term of Bayes' theorem \citep{Bayes1763}, which is is variously known as the ``prior predictive" \citep{Gelman2014}, ``marginal likelihood'' or ``evidence" \citep{Kass1995}:
\begin{equation}
    p(\theta|y) = \frac{p(y|\theta)p(\theta)}{p(y)} = \frac{p(y|\theta)p(\theta)}{\int_\theta p(y|\theta)p(\theta) d\theta}.
\end{equation}
Here the data is $y$ and the parameters are $\theta$, so $p(\theta|y)$ (read as ``the probability of theta given y") is the posterior, $p(y|\theta)$ is the likelihood, and $p(\theta)$ is the prior on theta.  The evidence $p(y)$ is a normalizing factor, so it is just the numerator integrated across theta.
 
Because the evidence does not depend on $\theta$, it is often left out of MCMC sampling as unnecessary.  However, it has found an application in model comparison, where the ratio of the evidences of the two models (say, $\mathcal{A}$ and $\mathcal{B}$) is known as the Bayes factor $B$
\begin{gather}
    B_\mathcal{AB} = B_\mathcal{BA}^{-1} = p(y|\mathcal{A}) / p(y|\mathcal{B}).
\end{gather}

Typically this is presented as with the greater evidence on top so that $B>1$. This quantity is valuable because it represents the ratio of probabilities of the models. To be precise, let $p(\mathcal{A}|y)$ and $p(\mathcal{B}|y)$ be the probabilities that each respective model is true given the data, the choice of prior, and that these two models are the only possibilities.  The last assumption means that the priors and posteriors both sum to one $p(\mathcal{A}) + p(\mathcal{B}) = 1$, $p(\mathcal{A}|y) + p(\mathcal{B}|y)$ = 1.  Applying Bayes rule again under equal prior probabilities $p(\mathcal{A})=p(\mathcal{B})$, we have \citep[following][]{Jeffreys1939, Kass1995}
\begin{equation}
    B_\mathcal{AB} \equiv \frac{p(y|\mathcal{A})}{p(y|\mathcal{B})} =
    \frac{p(\mathcal{A}|y)p(\mathcal{B})}{p(\mathcal{B}|y)p(\mathcal{A})} =
    \frac{p(\mathcal{A}|y)}{1 - p(\mathcal{A}|y)}.
\end{equation}
Solving for the probability of model A $p(\mathcal{A}|y)$, we find
\begin{equation}
        p(\mathcal{A}|y) = \frac{B_\mathcal{AB}}{1+B_\mathcal{AB}} = \frac{1}{1+B_\mathcal{BA}} \label{eq:bfInterpretation}
\end{equation}
So, a Bayes factor of 75 may be read as 75:1 odds in favor of the first model, given the aforementioned assumptions and choices. For the more general case of multiple models $\mathcal{M}_i=(\mathcal{M}_1,\mathcal{M}_2,\mathcal{M}_3...)$ being compared, the probability for model $i$ may be obtained through similar arguments as:
\begin{equation}
    p(\mathcal{M}_i|y) = \frac{p(y|\mathcal{M}_i)p(\mathcal{M}_i)}{\sum_j p(y|\mathcal{M_j})p(\mathcal{M}_j)}.
\end{equation}

A common approach for the interpretation of Bayes factors is via the Jeffreys table \citep[][see Table \ref{tab:interpretations}]{Jeffreys1935}.  This is a list of qualitative descriptions of the strength of evidence corresponding to various ranges of the Bayes factor.  For example $1<\log_{10}(B_{\mathcal{AB}})<1.5$ would be labeled as strong evidence for $\mathcal{A}$.  However, standards of evidence vary between fields and circumstance; astronomers would be unlikely to consider e.g. 15:1 odds in favor of a hypothesis to be strong statistical evidence for it.

It is important to note that these model probabilities are \emph{not} equivalent to p-values, which are instead defined as the fraction of potential outcomes under the null hypothesis which would be as or more extreme than the one observed.  On the basis that the interpretation of the Bayes factor is substantially more in line with researcher's intuitions, \citet{Sellke2001} present (and \citet{Trotta2008} repeat for astronomers) a formula for converting a p-value into a upper limit on the Bayes factor for $p<e^{-1}$,
\begin{equation} \label{eq:sellke}
    B \leq -\frac{1}{e p \ln(p)}.
\end{equation}

The main utility of this formula is to demonstrate the p-value fallacy directly by showing that the maximum Bayes factor derived from a given p-value \emph{excludes} $(1-p)/p$, the value it would take if the p-value represented a model probability.  \citet{Sellke2001} never discuss the possibility of solving for $p$ and in their conclusions expressly recommend \emph{against} the use of this formula for comparing two models.

Astronomers and physicists are accustomed to reporting their results as a significance $n_\sigma$, which relates to the p-value either directly or by metaphor to the two-sided Z-test where $n_\sigma$ is the z-score,
\begin{equation}\label{eq:sigmaDefinition}
    n_\sigma = -\Phi^{-1}(p/2).
\end{equation}
Here, $\Phi^{-1}$ is the inverse cumulative density function (CDF) of the normal distribution, also known as the percentile point function (PPF).  This formula is erroneously employed to p-values even when the test is not strictly a z-test and/or it isn't two-sided.

Since $n_\sigma$ has traditionally been calculated from a p-value, astronomers sought to calculate an equivalent p-value to the Bayes factors they computed, hence the use of Eq. \ref{eq:sellke}; we'll call this the inverse-Sellke method going forward.  Unfortunately this approach is not valid, despite having become popular among exoplanet atmosphere observers through a chain of miscommunications from \citet{Sellke2001} to \citet{Trotta2008} to \citet{Benneke2013} to the community at large (see Sec. \ref{sec:effects}).  This article seeks to rectify the misunderstanding.

In Sec. \ref{sec:problem}, we'll briefly cover the original \citet{Sellke2001} argument, observing that while it is valid, it doesn't remotely support the inverse-Sellke approach.  We'll show a simple example in which inverse-Sellke fails to bound several valid ways to derive p-values, though there is no exact analogue.  Generally, inverse-Sellke overestimates the significance.

We recommend researchers immediately switch to the standard Bayesian interpretation of their Bayes factors as an odds ratio or model probability via Eq. \ref{eq:bfInterpretation}.  However, this recommendation raises a second issue that bears mentioning due to the lack of well-motivated priors for exoplanet atmosphere models: the Bayes factors are strongly sensitive to the choice of priors.  Though this property is well known to statisticians, we believe the community underestimates the issue.  We'll discuss it in Sec. \ref{sec:priorSensitivity} with a brief example to motivate the use of information criteria in Sec. \ref{sec:recommendations}, with Python implementations in Appendix \ref{sec:python}.  We conclude with a case study of SO$_2$ in the atmosphere of WASP-39 b (Sec. \ref{sec:caseStudy}) and specific recommendations going forward.

\section{The Problem} \label{sec:problem}
The purpose of \citet{Sellke2001} was to relate the p-value from a simple hypothesis (meaning the null is fully specified; refer to \citet{Lehmann2022}, Ch. 3) to a Bayesian model probability.  This was accomplished by noting that the p-values of simple hypotheses are uniformly distributed from 0 to 1, and setting the Bayesian null model as being that the the observed p-value was drawn from this uniform distribution.  For a reasonable class of alternate models, they then identify a limiting alternate hypothesis which maximizes the Bayes factor for that p-value.  The resulting probabilities clearly rule out the erroneous interpretation of p-values directly as probabilities, a fact useful for combating some forms of the p-value fallacy.

As used in exoplanet astronomy however, we are starting with an existing null and alternate Bayesian hypothesis, neither of which matches the univariate setup originally presented.  Thus the limit provided by Eq. \ref{eq:sellke} \emph{does not apply}.  Moreover, the setup of the original argument takes as given that a statistical test with a simple hypothesis and p-value are already defined; going backwards, all of that is left undefined.  Because p-values are only meaningful in light the statistical test that produced them, the result is a non-bound on a meaningless number.

Recently, \citet{Kipping2025} disavow the inverse-Sellke approach as a misunderstanding of \citep{Benneke2013}, and that inverse-Sellke only gives a lower-bound on $p$ given $B$.  However, their reasoning is incorrect: the inequality in Eq. \ref{eq:sellke} points in the correct direction for how astronomers have been using it.  If $p$ places an upper limit on the significance of $B$, then knowing $B$ would indeed place a lower limit on the significance of $p$ (an upper limit on $p$) as $-p\ln(p)\leq1/(eB_{01})$\footnote{Notice that $-p\ln(p)$ is \emph{positive}, which may be the source of confusion.}.  The problem is instead that the assumptions of the relation are completely violated by the approach.

The arguments of \citet{Sellke2001} is useful to astronomers only for better understanding the p-values of simple hypotheses.  Eq. \ref{eq:sellke} is not applicable to the current approach for exoplanet atmosphere modeling, where we are starting from Bayesian models rather than p-values and rarely have simple hypotheses.

\subsection{Example}
\begin{figure}
    \centering
    \includegraphics[width=\columnwidth]{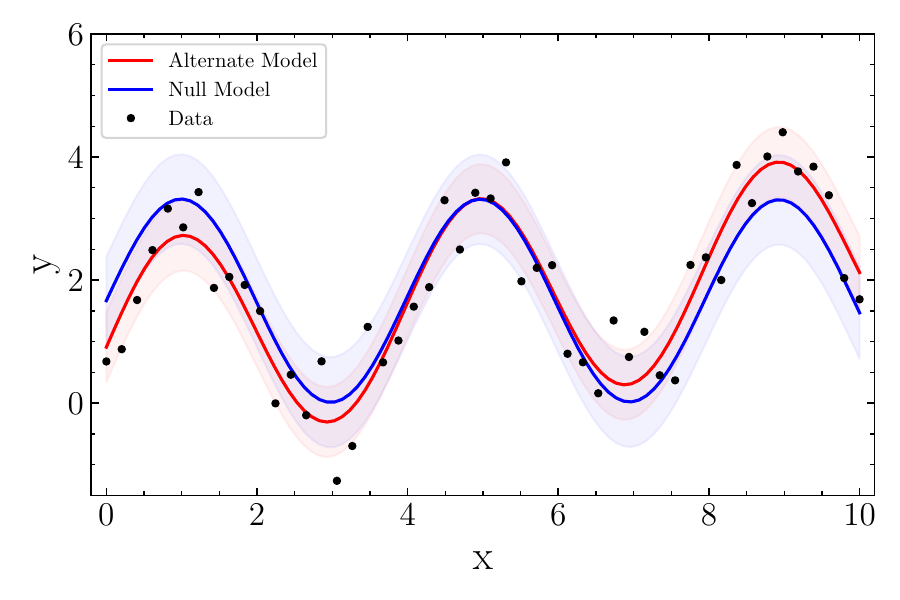}
    \caption{The data used in our example model comparison in black, along with the null (Eq. \ref{eq:nullModel}) and alternate (Eq. \ref{eq:alternateModel}) model posterior predictives (median and 16th to 84th percentile ranges).  The data were created from Eq. \ref{eq:exampleFunction} with $m=0.15$, so it was expected that the alternate model is preferred.}
    \label{fig:exampleCase}
\end{figure}

Although the invalidity of the the inverse-Sellke procedure has been established, an example of the procedure failing may be instructive.  For this, we will consider a simple case fitting the data shown in Fig. \ref{fig:exampleCase} with a line plus a sine wave:
\begin{equation}
    f(x, b, m, A, \lambda) = b + m x + A\sin(2\pi x/\lambda) \label{eq:exampleFunction}
\end{equation}
Here $x$ is known so we have three coefficients and the wavelength of the sine wave $\lambda$ as parameters.  We are interested in testing whether the slope term $m$ is non-zero.  Assuming the data $y$ are drawn from a normal distribution about $f$ with an unknown standard deviation $\sigma$, the null $\mathcal{A}$ and alternate $\mathcal{B}$ model posteriors are
\begin{align}
    p_\mathcal{A}(b, A,\sigma|y) \propto&\; \mathcal{N}(y|f(x, b, 0, A, \lambda), \sigma)\cdot \nonumber\\
        &\; p(b) \cdot p(A) \cdot p(\lambda)\cdot p(\sigma), \label{eq:nullModel} \\
    p_\mathcal{B}(b, m, A,\sigma|y) \propto&\; \mathcal{N}(y|f(x, b, m, A,\lambda), \sigma)\cdot \nonumber \\
        &\; p(b) \cdot p(m) \cdot p(A) \cdot p(\lambda)\cdot p(\sigma).\label{eq:alternateModel}
\end{align}

The only difference between the models is that the null has fixed the slope at zero.  For the priors, we'll set $b$, $m$ (omitted for the null), $A$, and $\lambda$ to be normally distributed with mean 0 and standard deviation 2, $\norm(0, 2)$.  For the error we'll set $\log(\sigma)\sim\norm(0, 2)$.  We'll talk more about these priors in Sec. \ref{sec:priorSensitivity}.  We draw $10^6$ samples from the distribution on 10 separate threads and verify convergence with the Gelman-Rubin statistic \citep{Gelman1992}.  We find the log-evidences given the data to be $\ln(p(y|\mathcal{A})) = -60.34$ and $\ln(p(y|\mathcal{B})) = -68.15$ for the alternate and null models respectively.  These result in a Bayes factor of $B_\mathcal{AB}=2465$, corresponding to a null probability $\probb=4.1\times 10^{-4}$.  The inverse-Sellke approach would yield $p_s^*=1.3\times 10^{-5}$ and $n_\sigma^*=4.36$.

While we'd like to calculate the p-value correctly to show that the inverse-Sellke method fails even as a limit, there is no directly analogous p-value to the Bayesian model comparison.  Instead, we'll calculate a few different statistics one might plausibly use in identifying an overall slope to the data and see that inverse-Sellke gives no constraint on any of them.  The Wald test on the correlation coefficient $r$ yields a test statistic of $r=.325$ and a p-value of $p_r=.0211$.  Similarly, Kendall's tau yields a test statistic of $\tau=.210$  and p-value of $p_\tau=.032$.  Both tests severely violate the supposed bound of $p_s^*$, exemplifying the meaninglessness of that calculation.  These tests are weakened by not accounting for the sin-wave, so we will consider some Bayesian tests that do next.

One approach to obtain a p-value within a Bayesian framework is to determine how probable a chosen test statistic as or more extreme than the one observed was under the prior predictive \citep{Box1980} or posterior \citep{Meng1994} predictive distributions.  That is,
\begin{equation}
    p_\pi = \int T(y,\theta)\pi(\theta) d\theta,
\end{equation}
where $\theta$ are the model parameters, $T(y, \theta)$ is the test statistic for $y$ and $\theta$, and $\pi(\theta)$ is the prior $p(\theta)$ or posterior $p(\theta|y)$ distribution depending on which version one wishes to compute.  Selecting Pearson's correlation coefficient $r$ (as before with the Wald test, so again $r=.325$) as the test statistic, we obtain a prior predictive p-value of $0.0318$ and a posterior predictive p-value of $1.6\times10^{-5}$.  Again, both clearly exceed the supposed limit, and only the posterior-predictive p-value is even close.  The results are similar using Kendall's tau as the test statistic.  Presumably one could engineer a scenario in which some p-value does not exceed the inverse-Sellke figure, but that would be artificial; the approach does not place any meaningful bounds on any p-value unless the setup exactly corresponds to the conditions of Sec. \ref{sec:problem}, but these are incompatible with the Bayesian models used for exoplanet atmospheres.

\section{Prior Sensitivity} \label{sec:priorSensitivity}
A second issue regarding the exoplanet community's use of the Bayes factor is its strong sensitivity to the prior.  This property of Bayes factors has been well known since the beginning \citep{Jeffreys1935}; it is desirable when priors are well motivated, as the Bayes factor incorporates the information in an statistically rigorous way.  For example, radial-velocity detection of exoplanets is an area where a great deal of thought has been put into the prior distributions \citep[e.g.][]{Ford2006a, Loredo2012, Nelson2020} so that Bayes factors may be employed confidently.  Unfortunately, for modeling the atmospheres of exoplanets we generally lack well-motivated priors for our parameters (see Sec. \ref{sec:caseStudy}).

In this situation, prior sensitivity is a serious concern.  It is part of why \citet[][section 7.4]{Gelman2014} recommends against the use of Bayes factors for model selection: ``the marginal likelihood is highly sensitive to aspects of the model that are typically assigned arbitrarily and are untestable from data".  Similarly the documentation for Dynesty \citep{Speagle2020, Koposov2024}, a sampling code commonly used in exoplanet science that returns Bayes factors, warns ``It cannot be stressed enough that the evidence is entirely dependent on the “size” of the prior."  As one expands the width of a prior into regions where the posterior is small, the evidence decreases as roughly the inverse of the width \emph{even if the posterior distribution is unchanged}.  What has been described as a natural or Occam penalty \citep{Kipping2013, Benneke2013} in the Bayes factor is only as well-motivated as the prior it is based on.

To illustrate this issue, consider again the models from the previous section (Eqs. \ref{eq:alternateModel} and \ref{eq:nullModel}), but instead set our prior to very weakly informative normal distributions, $\log(\sigma)\sim\norm(0, 1000)$ and $\norm(0, 1000)$ for the remaining parameters.  The resulting posteriors are largely identical, yet the Bayes factor has dropped precipitously to $B=9.89$ ($p_\mathcal{A} = .096$).  While most of our priors were identical between models, $p(m)$ was present in the alternate model but not in the prior, strongly penalizing the alternate model.  The issue compounds the more models differ; e.g. the addition of several chemical species at once or a species with several parameters (SO$_2$ in Sec. \ref{sec:caseStudy}) for exoplanet atmospheres.

This is especially important to understand for fitting models with nested sampling.  For this sampling method, a simple way to reach convergence faster is to narrow the priors; however, this inherently increases the evidence.  Even if applied uniformly across models and parameters, this decreases the penalty for adding parameters.  In the worst-case scenario, researchers may misreport their priors in the mistaken belief that if the resulting posteriors are the same the narrowed priors are equivalent to the original ones.  This could severely bias the Bayes factor.

One might imagine avoiding the issue through the use of improper priors (e.g. a ``flat" prior $p(a)\propto 1$), defined as any prior which cannot be normalized.  While these are very useful for obtaining less biased posteriors \citep[see examples in][]{Yang1996} so long as the data are sufficiently informative, they cause the model evidence integral to diverge (in fact, nested sampling \emph{requires} proper priors).  It is a fairly common practice to ignore this when the improper prior is present in and identical between all models considered \citep[][Sec 5.3]{Kass1995}; this could be interpreted as the limiting Bayes factor as the prior widths go to infinity.

All of this is not to say that Bayes factor analyses should not be employed; rather, researchers should be careful to test for prior-sensitivity in results and potentially augment their work with less prior-sensitive methods (see Sec. \ref{sec:recommendations})  Also, caution should be taken in using weakly-informative priors, as these may implicitly favor one model over the other as demonstrated.  Readers should refer to e.g. \citet{Kass1995, Trotta2008} for further discussion.

\section{Impact} \label{sec:effects}
\begin{figure}
    \centering
    \includegraphics[width=\columnwidth]{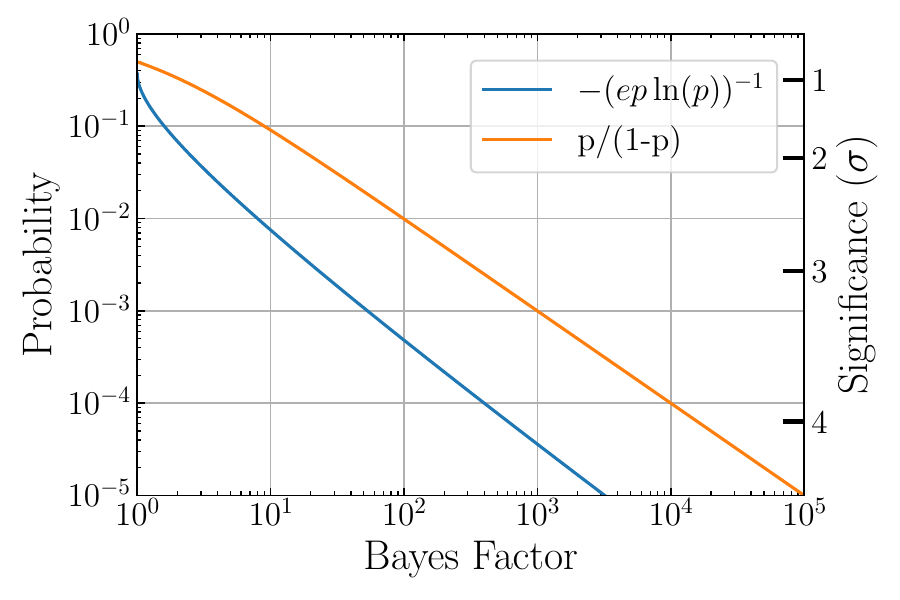}
    \caption{A comparison of the sigma associated with incorrect interpretation of \citet{Sellke2001} (right axis) with the model probability $\probb$ under the standard Bayesian interpretation (left axis) for a given Bayes factor $B$.  The erroneous procedure produces ``significant" values of sigma even for rather low odds ratios -- e.g. $3\sigma$ at 23:1 odds (probability $\probb=.042$) and $2\sigma$ at 2.6:1 odds ($\probb=.28$).}
    \label{fig:sigmaComparison}
\end{figure}

The difference between the invalid inversion of Eq. \ref{eq:sellke} and the Bayesian interpretation of the Bayes factor as a probability is substantial, as shown in Fig. \ref{fig:sigmaComparison}.  If Eq. \ref{eq:sellke} produced a valid p-value from a Bayes factor, it would be necessary to note that the interpretations are different (see introduction).  It isn't, so we are mainly concerned with how significant results are given the only valid interpretation left, Eq. \ref{eq:bfInterpretation}.  We will use the symbol $n_\sigma^*$ to denote reported results which used inverse-Sellke and give the proper Bayesian probability $\probb$ to replace it.

Searching through the exoplanet atmospheric observations literature, we have sought to identify how past results have been affected by this issue.  The use of Eq. \ref{eq:sellke} has been widespread, affecting nearly every exoplanet atmosphere paper using Bayes factors.  The exceptions are those which relied exclusively on the Jeffreys table without converting $B$ to $n_\sigma$, including \citet{Fisher2018, Seidel2020}.  Of course, papers which used non-Bayesian methods \citep[e.g.][]{Hoeijmakers2018, Moran2018} or Bayesian methods but not the Bayes factor \citep[e.g.][]{Benneke2012,Fraine2014} were unaffected entirely.  \citet{Kipping2013} (an exomoon search, not atmospheric spectroscopy) uses the Bayes factor with the correct formula for $p$, from which he calculates $n_\sigma$ via Eq. \ref{eq:sigmaDefinition}.

A number of molecular detections were so strongly significant that even correcting their significances down does not impact their conclusions.  Such cases include the detection of CO$_2$\citep{jwstCollab2023}, water and SO$_2$ \citep{Alderson2023} in WASP-39 b, of water in K2-18 b by \citet{Benneke2019}, of water in WASP-12 b \citep{Kreidberg2015a}, and quite a few others \citep[e.g.][]{Line2016a, MacDonald2017, Zhang2021a, Chen2021, Fu2024, Rotman2025, McCreery2025}.  In \citet{Bell2023}, the H$_2$O in emission was non-significant $\nss=2.59$, but the detection was stronger in transmission ($\nss=4.6\sigma$ to $\probb=1.4\times10^{-4}$) so it hardly matters.  Likewise, papers which use the formula to determine non-significance or tentative detections are fine \citet{Mikal-Evans2021, Moran2023, Gressier2024}, as inverse-Sellke overestimates the significance.

Some results compute $n_\sigma$ as an aside in conjunction with other model comparison methods, such as explicitly invoking the Jeffreys table as the main determinant \citep[e.g.][]{Feng2016, Guilluy2021, Brande2022} or using the Bayes factor in conjunction with other tests like the BIC, AIC or cross-validation \citep[including][]{Alderson2022, Welbanks2023, Murphy2024,Alderson2023}.  Similarly, \citet{May2023} uses arguments based on the $\chi^2$ goodness-of-fit in addition to $n_\sigma$.

Most affected were results in the 2-3.5$\sigma$ range, whose significance now depends on where one chooses to place the significance threshold.  A recent high-profile result in this category is the detection of dimethyl sulfide (DMS) or dimethyl disulfide (DMDS) on K2-18 b \cite{Madhusudhan2025} -- the detection significance of a non-flat atmosphere is reduced from $\nss=3.4$ to $\probb=.013$ and of DMS/DMDS from $\nss=3$ to $\probb=.042$.  Other works on K2-18 b are similarly affected, including \citet{Tsiaras2019} and \citet{Madhusudhan2020}, though the detection of an atmosphere containing water were statistically significant in other works \citep[e.g.][]{Benneke2019, Madhusudhan2025}.

Other detections in the $2 < \nss < 3.5$ range include water on WASP-117 b \citep{Carone2021}, some (but not all) of the dataset analyses of \citet{Swain2014}, the various molecules reported in LHS 1140 b \citep{Damiano2024}, the detection of MgH in \citep{Kang2024}.  The detection of SO$_2$ on WASP-39 b by \citet{Rustamkulov2023} was another such case, and we reexamine it in section \ref{sec:caseStudy}.

Finally, the use of inverse-Sellke appears to have recently spread to the brown dwarf observations community as well \citep[e.g.][]{Xuan2022, Adams2023, DeRegt2024, Nasedkin2024, Grasser2025, Mulder2025, Kuhnle2025}.  Ultimately, it is not possible to get a complete list of affected papers as many report only a $n_\sigma$ and not the Bayes factor itself \citep[e.g.][]{Welbanks2019, Kempton2023a, Taylor2023}; we can't conclusively determine which formulas for $p$ and $n_\sigma$ were used.  The lack of reporting sufficient statistical information for readers to independently verify and reproduce results is a recurrent issue that should be addressed by the community, as it contributes to the broader replication crisis in science.

Perhaps the most important instances of this error are in the planning of future observations.  It is a long-standing practice in astronomy to use mock retrievals to simulate telescope observations and justify the needed time. The detection threshold is traditionally set to $n_\sigma=3$, so the use of Eq. \ref{eq:sellke} underestimates the required observation time.  Telescope time allocated on this basis would be highly susceptible to non-significant results and weak detections. How widespread this issue is cannot be determined as telescope proposals are not public and rarely detail such in-depth statistical methods.

Finally, we note there are other additional statistical methodological issues that may be common in the field which may further affect the confidence of reported results. For instance, lack of reporting and accounting for correlated (red) noise , insufficient reporting of convergence diagnostics, and not accounting for data reduction differences in the uncertainties. For the latter issue, when comparing the same dataset agreement much closer than 1 or 2 $n_\sigma$ is expected between different pipelines given the identical underlying data, with  disagreements pointing to underestimated or unaccounted for systematic uncertainties. 

\section{Recommendations} \label{sec:recommendations}
While we would like to recommend a single new approach to model selection, there is no one method that is clearly superior and can be applied in every situation.  Exoplanet atmosphere retrievals are particularly difficult to compare due to several details of the situation.

First, the binning is nearly arbitrary -- red noise means declining to bin the data underestimates uncertainty \citep{Pont2006}, yet the issue is not well enough understood that we can just set a covariance or use a clear rule for binning.  Additionally, the wavelength cutoffs used (e.g. to avoid saturation) are largely arbitrary.  These issues can lead to the numbers of data points varying between different reductions of the same data.  Using the full spectral range of the detector and avoiding binning can potentially alleviate these issues, as the number of detector pixels is fixed and a known quantity. However in practice issues such as partial saturation of a region of the detector and low-level systematics which require binning to adequately model in the light curves often inhibits this practice.

Second, many of the parameters lack well-motivated priors; the abundances in particular are often modeled with log-uniform priors (see Sec. \ref{sec:caseStudy}) but it is unclear what the lower limit should be, if any.  Further, the inclusion or exclusion of molecules from consideration affects the implied prior (ultimately, volume mixing ratios must add to one), but it is not well-established which molecules deserve consideration and with what prior weight.  Of course, this likely would vary with the type of planet, complicating matters further.

Third, the forward models are very computationally expensive and often of high enough dimension that pre-gridding models is infeasible.  One simplifying factor is that recent Bayesian models of exoplanet atmospheres have rarely been hierarchical; however, future models might benefit from such structures.  We'll consider several model comparison alternatives to the Bayes factor in light of these complications.

Note that we'll be describing these in terms of the likelihood, but as the chi-squared statistic $\chi^2$ (known as the deviance to statisticians) may be more familiar to some readers, we'll note their relation for normally-distributed independent errors here \citep{Bevington1992}.
\begin{align}
    \chi^2 &= \sum_i \left( \frac{y_i-\mu_i}{\sigma_i} \right)^2\\
    \ln(L) &= -\chi^2/2 - \sum_i\ln(2\pi\sigma_i^2)/2 \label{eq:loglikechi}
\end{align}
Here, $L$ is the likelihood, $y_i$ is the $i^\text{th}$ datapoint, $\mu_i$ is the corresponding predicted value, and $\sigma_i$ is the error on it.  In comparing two log-likelihoods on the same data, the second term of Eq. \ref{eq:loglikechi} cancels out.  Be aware that this cancellation occurs exclusively the case if $\sigma_i$ are constant and independent; for example, if $\sigma_i$ are determined hierarchically the terms do not cancel.  Also note that $\chi^2$ is \emph{not} the statistic from Pearson's chi-squared test, which relates to categorical distributions.

\begin{table*}
    \centering
    \begin{tabular}{rccccccc}
        Jeffreys & $\log_{10}(B)$ & $\ln(B)$ & $B$ & $p(\mathcal{A})$ (\%) & $p(\mathcal{B})$ (\%) & $\Delta$AIC & Usage \\
        \hline
        \ldelim\{{1.9}[-5pt]{*}[Barely Worth Mentioning]   &  0.00 &  0.00 &        1.0 & 50.00         & 50.000              &   0.00 &                     \\ 
        \ldelim\{{0.9}[-20pt]{*}[Substantial]              &  0.33 &  0.77 &        2.2 & 68.27         & 31.731              &  -1.53 &                     \\ 
        \ldelim\{{1.85}[-20pt]{*}[Strong]                  &  0.50 &  1.15 &        3.2 & 75.97         & 24.025              &  -2.30 &                     \\ 
        \ldelim\{{0.9}[-35pt]{*}[Very Strong]              &  1.00 &  2.30 &       10.0 & 90.91         & 9.091               &  -4.61 &         Indications \\ 
        \ldelim\{{2.9}[-36pt]{*}[ Decisive]                &  1.32 &  3.04 &       21.0 & 95.45         & 4.550               &  -6.09 &       Minor Results \\ 
                                                           &  1.50 &  3.45 &       31.6 & 96.93         & 3.065               &  -6.91 &                     \\ 
                                                           &  2.00 &  4.61 &      100.0 & 99.01         & 0.990               &  -9.21 &                     \\ 
                                                           &  2.57 &  5.91 &      369.4 & 99.73         & 0.270               & -11.82 &          Detections \\ 
                                                           &  4.20 &  9.67 &    15786.2 & 99.9936   & 0.006               & -19.33 &                     \\ 
                                                           &  6.24 & 14.37 &  1744276.9 & 99.999943   & $5.73\times10^{-5}$ & -28.74 &   Major Discoveries \\
        \hline
    \end{tabular}\hspace{5cm}
    \caption{A comparison of various approaches to interpreting statistical tests related to the Bayes factor.  The Jeffreys table \citep[][appendix]{Jeffreys1939} is represented in the first column next to various transforms of the Bayes factor $B$.   The corresponding probability $p$ is the next column \ref{eq:bfInterpretation}, followed by an analogous $\Delta \mathrm{AIC}$ following \citep{Akaike1978}.  Finally we suggest \emph{rough} significance targets for various types of results on the right; these should not be used as hard cutoffs.}
    \label{tab:interpretations}
\end{table*}

\subsection{Comparisons from the Maximum Likelihood}
\subsubsection{BIC}
Perhaps the best-known model selection criterion is the Bayesian Information Criterion \citep{Schwarz1978}, which was constructed as a prior-free approximation to the Bayes Factor.  It is well-established in statistics generally \citep{Gelman2014} and in exoplanet science \citep[e.g.][]{Gibson2010, Sing2011, Mandell2013, Thorngren2021}.  It is defined as the model deviance plus a penalty for adding parameters:
\begin{align}
    \text{BIC} &= k\ln(n) - 2\ln(\hat{L}) \label{eq:BIC}\\
    &= \chi^2 + k\ln(n) + c \label{eq:BICChi}
\end{align}
In these equations, $k$ is the number of parameters, $n$ is the number of data points, and $\hat{L}$ is the maximum likelihood (remember to exclude priors from this value).  The second form using $\chi^2$ assumes independent normal errors; $c$ is a constant that depends only on the data (see Eq. \ref{eq:loglikechi}), so it is the same between models and cancels out.  To use the BIC, one must compute the statistic for each model; the lesser model is preferred.

A useful way to understand the BIC is as "a rough approximation to the logarithm of the Bayes factor" \citep{Kass1995} with prior information excluded.  Following \citet{Konishi2008}, the integral of the evidence is approximated using the Laplace approximation, a second-order series expansion around the mode $\hat\theta$.  This amounts to approximating the posterior as multivariate normal with mean $\hat\theta$ and covariance $n\mathcal{J}=-\nabla^2p(y|\hat\theta)$.  $\mathcal{J}$ is the average observed information, defined as the Hessian of the negative log-likelihood at the mode $\hat\theta$, divided by $n$.
\begin{align}
    &p(y) = \int p(y|\theta) p(y) d\theta \\
         &\approx \int p(y|\hat\theta) p(\hat\theta) \exp\left[ -\frac{n}{2}(\theta-\hat\theta)^\mathrm{T} \mathcal{J}(\hat\theta)(\theta-\hat\theta)\right] d\theta\\
         &\approx p(y|\hat\theta) p(\hat\theta) (2\pi/n)^{k/2} \left|\mathcal{J}(\hat\theta)\right|^{-1/2} \\
         &\approx \exp\left(-\mathrm{BIC}/2\right) p(\hat\theta) (2\pi)^{k/2} \left|\mathcal{J}(\hat\theta)\right|^{-1/2} \label{eq:BICDerivation}
\end{align}
Thus in addition to the assumption of multivariate-normality, the BIC also excludes the three terms in the right side of equation \ref{eq:BICDerivation}.  Excluding $p(\theta)$ amounts to ignoring prior information, which may be desirable if priors are poorly-motivated; excluding $(2\pi)^{k/2}$ causes the BIC to systematically favor simpler models.  Excluding the $\mathcal{J}$ term is roughly equivalent to assuming each data point constrains the posterior covariance similarly across models.  On the other hand, for nested hypotheses \citet{Kass1995a} argues that the assumptions amount to a reasonable reference prior based on $\mathcal{J}$ (as well as some regularity assumptions) which make the BIC an O($n^{1/2}$) approximation to the Bayes factor rather than the usual O(1).  For the general case, the approximation is indeed rather rough.

The BIC has several drawbacks which make it difficult to recommend for the specific case of atmospheric retrieval comparison.  First, it relies only on the maximum likelihood, which may not be a sufficient summary of the posterior distributions, which often deviate substantially from a multivariate normal.  Second, it scales its penalty as $\ln(n)$, but in this context the number of data points can change with different choices of spectral binning (see discussion in Sec. \ref{sec:recommendations}).  Finally, a penalty derived from the number of data points across an entire broad spectra seems undesirable when considering adding a molecular abundance which only affects a narrow wavelength span.

\subsubsection{AIC}
A similar model selection criterion is the AIC \citep{Akaike1974}, which differs from the BIC only in its choice of penalty,
\begin{align}
    \text{AIC} &= 2k - 2\ln(\hat{L}) \label{eq:AIC}\\
    &= \chi^2 + 2k + c \label{eq:AICChi}
\end{align}
The variables are all defined the same as for the BIC.  The change means the AIC is more favorable towards additional parameters when $n\geq8$, and that it no longer matters how many data points are used.  The difference in penalty is the result of a different objective: the AIC is based on an information theory argument rather than the Bayes factor.  That said, its creator has argued that it can be interpreted similarly to the BIC as a prior-excluded approximation to the Bayes factor up to a constant $C$ \citep{Akaike1978,Akaike1981}
\begin{equation} \label{eq:aicScaling}
    p(y) \approx C \exp(-\Delta\mathrm{AIC}/2),
\end{equation}
a relation that has seen use in exoplanet astronomy \citep[e.g.][]{DelGenio2019,Tran2022} and other fields \citep[e.g.][]{Bozdogan1987, Anderson1998, Burnham2011}.  This approximation is quite rough, perhaps even more so than the BIC's equivalent; if a Bayes factor is desired one should compute a Bayes factor.  Still, the formula is useful for understanding how the $\Delta$AIC is scaled.  Overall, we find the AIC to be preferable to the BIC for atmospheric retrievals because the binning issue is bypassed, but the fact that it is a point estimate remains an issue.

\subsection{Comparisons from Samples}
\subsubsection{WAIC}
Cross-validation is a well-established and robust technique for model evaluation and selection.  Refitting each model on subsets of the data can be computationally expensive, however, so approximations based on a single fitting run can be helpful.  The widely-applicable information criterion \citep[WAIC, ][]{Watanabe2010} fills this role as an approximation to leave-one-out cross-validation which can be computed from a single MCMC run. It requires the pointwise likelihood of $n$ independent data points to compute, which can be preserved in Emcee and Dynesty via the ``blobs" feature; see the example in the documentation of our implementation in Appendix \ref{sec:python}.
\begin{align} \label{eq:waic}
    \text{WAIC} = -2 &\sum_i \ln\left(\text{E}_\theta [p(y_i|\theta_s)]\right) + \\
                2 &\sum_i \text{Var}_\theta(\ln(p(y_i|\theta)) \nonumber
\end{align}
That is, $-2$ times the log of the mean likelihood minus the variance of the log likelihood, summed across points.  Here, $i$ indexes the data point and E$_\theta$ and Var$_\theta$ are the mean and variance across posterior samples.  Note that the first term is the ``elpd$_\text{loo}$" used in \citet{Welbanks2023}, though they do not consider the subsequent penalty term found in the WAIC \citep{Watanabe2010, Vehtari2017}.

For all its benefits, the WAIC isn't well-suited for exoplanet retrievals due to the arbitrary binning / correlated noise issue.  If it is used, it should be tested with multiple different binning schemes to ensure the choice was not important and the penalty term should be included.

\subsubsection{DIC}
The DIC \citep[Deviance Information Criterion, ][]{Spiegelhalter2002} is a simpler alternative for comparing models with sampled posteriors \citep[used in e.g.][]{Jordan2013, Chen2016, Thorngren2018, Morello2019}.  It does not require pointwise likelihoods like the WAIC and is easy to compute without any modifications to the MCMC procedure.  The original formulation was criticized for under-penalizing extra parameters \citep{VanDerLinde2005,Ando2011,Spiegelhalter2014}, so using the increased penalty suggested by \citet{VanDerLinde2005}, we have
\begin{align}\label{eq:DIC}
    \text{DIC} &= -2\ln(p(y|\bar{\theta})) + 3P_D \\
    P_{D1} &= 2\ln(p(y|\bar{\theta}))) - 2 E_\theta[\ln(p(y|\theta))] \label{eq:effectiveParams1} \\
    P_{D2} &= 2\textrm{Var}_\theta[\ln(p(y|\theta))] \label{eq:effectiveParam2}
\end{align}
The term $\ln(p(y|\bar{\theta})$ is the likelihood at the posterior mean of the parameters, $E_\theta[\ln(p(y|\theta))$ is the posterior mean of the likelihood, $\text{Var}_\theta[\ln(p(y|\theta))$ is the posterior variance of the likelihood, and $P_D$ is the effective number of parameters.  Two formulas for $P_D$ are given: $P_{D1}$ is the original approach of \citep{Spiegelhalter2002}, and $P_{D2}$ is from \citet{Gelman2014}, chosen to be invariant to the parameterization and make the number of parameters strictly positive -- the former approach can yield a negative $P_D$ if the mean is too far from the mode, a good example of the drawbacks of using point-estimates on complex distributions.  This second method is always positive, but comes at the cost of some numerical stability, so be careful to use an especially large number of posterior samples in its estimation.  While the DIC is an improvement for using the posterior information, it still has the weakness of depending on a point estimate at $\bar{\theta}$ for both versions of $P_D$.

\subsubsection{BPICS} \label{sec:bpics}
An underappreciated criterion related to the DIC was proposed by \citet{Ando2011} as a simplification of their somewhat impractical BPIC \citep[Bayesian Predictive Information Criterion;][]{Ando2007}.  The criterion was not given an acronym, but was described as a simplified BPIC, so we will refer to it as the BPICS.  Like the AIC and WAIC, this was constructed to select the model which has the best predictive power:
\begin{equation} \label{eq:bpics}
    \textrm{BPICS} = -2E_\theta[\ln(p(y|\theta)] + 2P_D.
\end{equation}
The variables here are defined in the same manner as with the DIC -- in fact, the BPICS could be viewed as an alternate form of the DIC, though \citet{Ando2011} derives it from first principles.  The $\Delta\text{BPICS}$ is scaled the same as the AIC, so Eq. \ref{eq:aicScaling} can be used as a guide for interpretation. While one may use the effective number of parameter estimates from Eqs. \ref{eq:effectiveParams1} and \ref{eq:effectiveParam2}, Ando suggests that one may optionally use the actual number of parameters used in the model, $P_D=k$.  While this requires some additional regularity assumptions to justify \citep[again see][]{Ando2011}, the practical benefits are substantial.

This form of the BPICS doesn't rely on a point estimate, is numerically stable, is insensitive to the priors (though not completely invariant), is insensitive to binning choices, and is both intuitive and easy to use.  It may be calculated directly from the posterior likelihoods, which most MCMC samplers save.  The penalty term is identical to the AIC, though note that the fit term also penalizes extra parameters somewhat -- at higher dimension, a larger portion of the posterior mass will be at lower probabilities.  For all of these reasons, we feel that the BPICS is the best information criterion available for comparing exoplanet atmosphere models.

\section{Case Study}\label{sec:caseStudy}
\begin{figure*}
    \centering
    \includegraphics[width=.9\textwidth]{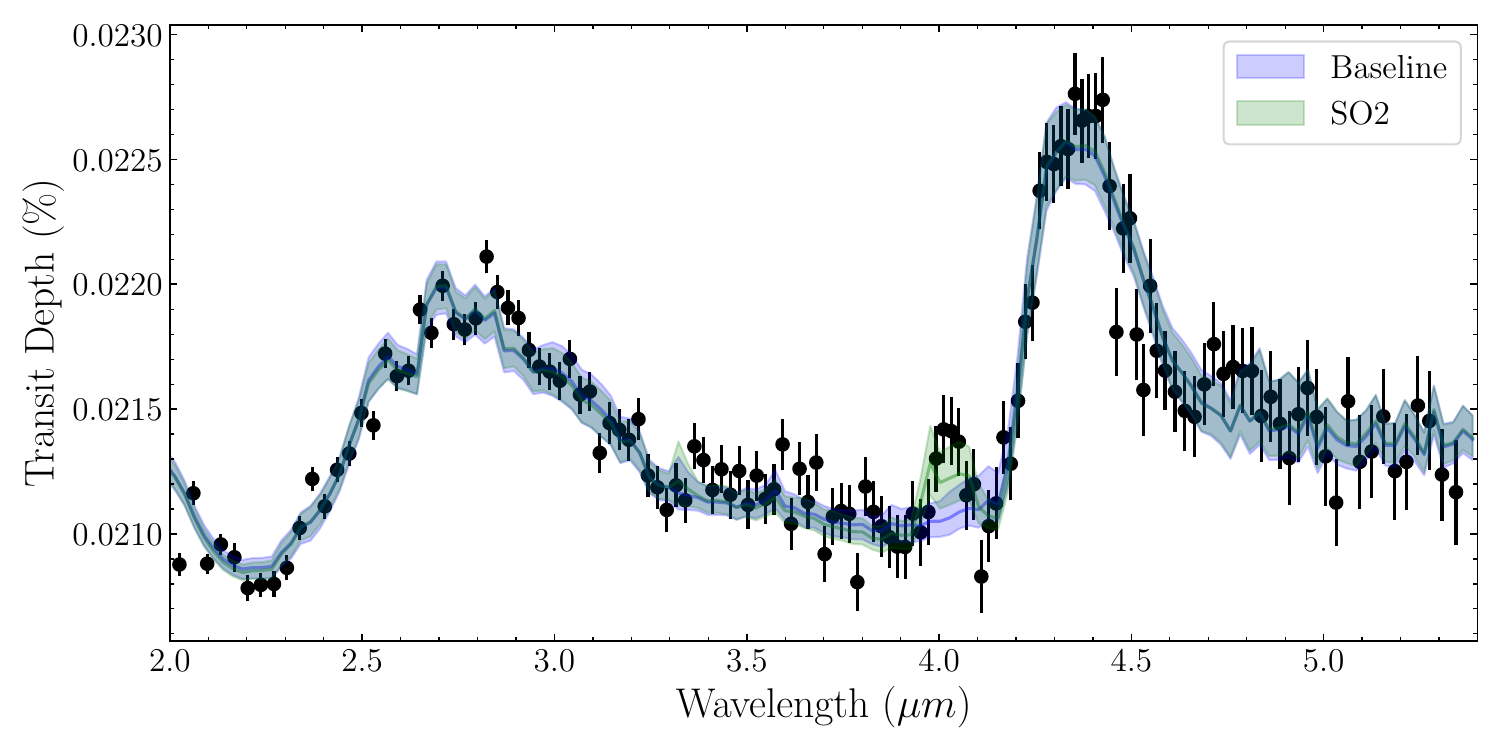}
    \caption{The WASP-39 b transmission spectrum captured by JWST NIRSpec and reduced in \citet{Rustamkulov2023}, with our model fits overplotted as the medians and $1\sigma$ contours.  Two models are given, one with sulfur dioxide and one without.  The SO$_2$ model is moderately favored by both the BIC, DIC, and Bayes factor, but does not exceed $3\sigma$.  Inverting Eq. \ref{eq:sellke} erroneously gives a significance above $3\sigma$.}
    \label{fig:spectrum}
\end{figure*}
\subsection{WASP-39 b SO$_2$ Retrieval}\label{sec:retrieval}
To demonstrate how the different approaches, especially inverse-Sellke, affect the outcome of model comparison, we have fitted two variants of a 1D atmosphere model to the NIRSpec/PRISM transmission spectrum of WASP-39 b from \citet{Rustamkulov2023} with wavelengths greater than 2\micron. We aim to re-evaluate the detection of SO$_2$ gas in WASP-39 b's transmission spectrum with our statistical framework. The choice of SO$_2$ in WASP-39 b is motivated by two factors: first, the original evidence from the NIRSpec/PRISM data was near the $3\sigma$ detection threshold; and second, the molecule represents an important discovery - providing the first substantial evidence for exoplanet photochemistry \citep{Tsai2023}. We limited the study to the data long-wards of 2{\micron} to avoid issues with saturation at shorter wavelengths.

We have used the \texttt{PICASO} 1D atmosphere model \citep{Batalha2019,Mukherjee2023} to perform a Bayesian retrieval analysis on the observed transmission spectrum of WASP-39 b. We divide the model atmosphere into 100 plane-parallel layers that are logarithmically spaced in pressure between 10$^{-9}$-1 bars. We use the parameterization from \citet{Guillot2010} to model the temperature--pressure ($T(P)$) structure of the atmosphere with five input parameters-- $T_{\rm eq}$, $log(g)$, $log(\kappa_{\rm IR})$, $\alpha$, and $T_{\rm int}$. We include vertically uniform volume mixing ratio of gasses H$_2$O, CH$_4$, CO$_2$, CO, NH$_3$, H$_2$S, HCN, C$_2$H$_2$, OCS, C$_2$H$_4$, and PH$_3$ as parameters within the model. We also include a vertically non-uniform volume mixing ratio profile for SO$_2$, as it can be predominantly produced within a narrow pressure-range of the atmosphere due to photochemistry \citep{Tsai2023,Mukherjee2024}. We model the volume mixing ratio profile for SO$_2$ using three free parameters $P_{\rm 1}$, $P_{\rm 2}$, and $X_{\rm SO_2}$. The volume mixing ratio profile for SO$_2$ is given by,
\[
X_{\rm SO_2} (P) = 
\begin{cases}
  X_{\rm SO_2} & \text{if $P_1<P<P_2$} \\
  0 & \text{otherwise}
\end{cases}
\]
where $P$ is the pressure. The mixing ratio of H$_2$ and He are calculated by subtracting the mixing ratio profiles of these 14 gases from 1 and partitioning the remaining mixing ratio between H$_2$ and He assuming a H$_2$/He ratio of 5.134. In addition to these 14 chemical composition related parameters, we also include gray aerosol opacity in our model with the $\kappa_{\rm cld}$ parameter. The gray aerosol optical depth along the vertical direction of each atmospheric layer is given by,
\begin{equation}
    \tau_{\rm cld}= \dfrac{\kappa_{\rm cld}\delta{P}}{g}
\end{equation}
where $\delta{P}$ and $g$ are the pressure difference between adjacent atmospheric layers and $g$ is the gravity of the planet. We also include a $x_{R_p}$ parameter that scales the planet radius at 1 mbar as $R_{1 mbar}=R_P(1+x_{R_p})$. Therefore, our complete atmospheric model has a total of 21 free parameters. We assume system parameters including planet radius ($R_p$=1.27$R_{\rm J}$) and stellar radius ($R_*$=0.939$R_{\odot}$) from \citet{Mancini2018}. The planet mass was assumed to be $M_{\rm p}=0.281 M_{\rm J}$ from \citet{Mancini2018} as well. The source of the used opacity line-lists for the included gasses are described in \citet{Mukherjee2024a}.

We apply uniform priors on all the 21 parameters of the model. The logarithm (base 10) of the volume mixing ratio of the 15 gasses have been allowed to vary between -1 and -11. The $log_{10}(\kappa_{\rm cld})$ parameter is allowed to vary between -3 to 2. $log_{10}(P_1)$ and $log_{10}(P_2)$ are allowed to vary between -6 and +1 (in bars). $x_{R_p}$ is allowed to vary between -0.1 and +0.1. The $T_{\rm eq}$ was varied between 400 to 1900 K, $log(g)$ between -2 and +0.5, $log(\kappa_{IR})$ between -2 and +0.5, $\alpha$ between 0 and 1, and $T_{\rm int}$ between 30 K and 630 K. We use the PyMultiNest nested Bayesian sampler \citep{Buchner2014} with 2000 live points to sample the posterior these 21 parameters.

In addition to running this complete model, we also run a retrieval model without SO$_2$. We omit the three SO$_2$ volume mixing ratio-related parameters $X_{\rm SO_2}$, $P_1$, and $P_2$ from this model keeping everything else identical between this and the complete model.

\subsection{WASP-39 b SO$_2$ Significance Comparisons}
The results of these retrievals are shown in Fig. \ref{fig:spectrum}, along with the data they were fit to.  The model featuring SO$_2$ is better able to fit the feature at $\sim4$\micron, but with only four data points in that region, it isn't clear by inspection whether the difference is significant.  The Bayes factor between the models is 221:1 in favor of the complete model including SO$_2$.  Using the inverse-Sellkie approach yields $n_\sigma^*=3.73$, above the traditional cutoff of 3 for statistical significance; however, the actual model probability $\probb=0.0044$ isn't as clearly significant.  Thus this is an example where using the correct formula changes the qualitative result.  Making matters worse, we did not have well-motivated priors for the SO$_2$ model parameters, adding ambiguity to the outcome -- different plausible priors could have yielded a substantially different Bayes factor.

It is therefore helpful to support this approach with other model selection criteria.  First, the reduced $\chi^2$ of the models were 1.89 for the SO$_2$ model (with 116 DOF), and 2.008 for the baseline model (with 119 DOF).  These reduced $\chi^2$ values are high, largely due to several outliers especially a 7-$\sigma$ outlier affected by saturation just long-ward of 2 microns. However, the region near the SO$_2$ feature is well fit by the atmospheric model.  The unreduced chi-squared statistic were $216.85$ with SO$_2$ and $236.94$ without.  Adding a penalty for the number of parameters (see Eq. \ref{eq:AIC}), we get a significant $\Delta \mathrm{AIC}=-14.09$ (see Table \ref{tab:interpretations}, $\Delta$AIC column).  If instead we calculate the BIC, we get a far weaker $\Delta \textrm{BIC}=-5.35$ ($\probb \approx .064$).  We do not trust this value as the penalty is much higher than the AIC due to the number of points, yet the number of data points was the result of arbitrary binning and bounds choices.  For this reason and the approximations discussed section \ref{sec:recommendations}, the transformed BIC was indeed only roughly similar to the Bayes factor (though the AIC would be worse).

We have sampled the posterior distribution, so it makes sense to use those results in our model comparison.  Our number of data points is poorly defined, so the WAIC isn't well-defined here, but the DIC and BPICS still are.  For the former, we find that $\Delta DIC = -2.20$, favored but definitely non-significant.  We distrust this result as well, because the number of parameters estimate was \emph{negative} for $P_{D1}$ due to the likelihood at the posterior mean of the parameters being offset from the mode, which was in turn the result of a skew distribution in several parameters.  For our DIC calculation, we used $P_{D2}$ instead, which gave the effective parameter count at 19.49 with SO$_2$ and 13.87 without.  These numbers are at least plausible, but because the DIC itself uses the likelihood at the posterior mean (Eq. \ref{eq:DIC}), we do not feel we can trust the overall result; it will be too sensitive to the skewness of the parameter posteriors.

The BPICS has none of the aforementioned issues if we use $P_D=k$.  The fit term is the average log-likelihood so posterior skewness is a non-issue, and the penalty doesn't scale with the (arbitrary) number of data points.  We find that $\Delta BPICS=-13.07$ (again see Table \ref{tab:interpretations}).  This is similar to the result of the AIC, which is reasonable since both methods are designed to favor the model with the best predictive power.

Overall, we feel that the BPICS and AIC are the best model selection criteria to use followed by the Bayes factor, whereas the BIC, DIC, and WAIC are unfortunately poor fits for the exoplanet spectral fitting use-case.  The BPICS has the advantage of a slightly stronger penalty than the AIC (for subtle reasons, see Sec. \ref{sec:bpics}), and of using the whole posterior rather than the maximum likelihood.

As for SO$_2$ on WASP-39 b using the PRISM data, we strongly favor the presence of the gas in the planet's atmosphere, but cannot unambiguously establish that the result has $n_\sigma > 3\sigma$.  Subsequent studies have more firmly established the presence of SO$_2$ on the planet using additional observations \citep{Tsai2023,Alderson2023,Powell2024}.

\section{Conclusions} \label{sec:conclusions}
We have shown that the inverse-Sellke approach is not a valid way to compute a p-value, being both proscribed for two-model comparison in its original source \cite{Sellke2001} and completely diverging from the modeling setup of the inequality's derivation.  We recommend immediately discontinuing use of this formula and using the Bayesian interpretation as an odds ratio or the corresponding probability (Eq. \ref{eq:bfInterpretation}) in its place.  If a p-value is required, it is necessary to instead conduct an appropriate frequentist test rather than attempt to convert from an existing Bayesian approach.  It is especially important that mock-retrievals used to justify the required telescope time for observations revise their significance calculations, lest they underestimate the required time to achieve their target significance.

Due to the particular challenges of exoplanet spectra -- data correlations, the practice of spectral binning, saturation issues, irregular posteriors, poorly-motivated priors, etc. -- we recommend supplementing Bayes factor model comparisons with the use of the AIC \citep{Akaike1974} or the BPICS \citep{Ando2011} for comparison using the full MCMC-sampled posteriors.  For other situations where the number of data points does depend on data reduction or analysis choices, the WAIC is a strong option.  These should be reported with the method used and the statistic, e.g. ``the BPICS favors model A over B with $\Delta$BPICS=-23.0". Regardless of the main model selection approach, analyses should include the chi-squared statistic (including the number of degrees-of-freedom) to understand goodness-of-fit and improve reproducability.

These challenges also highlight the importance of carefully considering the priors to be applied, as seen in related subfields \citep[e.g.]{Ford2006a}.  Future such work applied to exoplanet atmospheric modeling would be valuable.  Similarly, a better understanding of red noise in JWST spectra would greatly improve our ability to interpret them accurately.  This should become increasingly feasible as more observations are made.

We have listed several equivalent ways to represent Bayesian statistical results in Table \ref{tab:interpretations}.  These include the Jeffreys table of \citet{Jeffreys1939}; however, we find that the descriptions are more generous than how astronomers would usually interpret results \citep[see related discussion in][]{Nesseris2013}.  For example, 10-to-1 odds is listed as the start of ``strong" results, but Astronomers typically require $n_\sigma \geq 3$ for a statistically significant ``detection''.  While there is no true translation from Bayes factor to sigma, the Jeffreys table clearly implies a lower standard of proof.

We feel that hard significance thresholds are undesirable; strong thresholds for publishability create perverse incentives for scientists and disregard probable but non-significant results. For example, results near 2.5$\sigma$ are unlikely to be statistical noise, but could still benefit from follow-up observations. It would be better to understand our results as a gradient in confidence in our conclusions (given the modeling assumptions, which must be considered separately).

With that in mind, we have listed \emph{rough} guidelines for how strong a result should be in the final column of Table \ref{tab:interpretations}, premised on the notion that highly-impactful results should have a higher standard of evidence.  The discovery of life and the presence of an atmosphere on a planet expected to have one should not require the same statistical confidence.  Researchers must also remain mindful of systematic errors in the data \citep[e.g.][]{Pont2006}, which have long hindered the exoplanet field by artificially inflating the confidence of results. While JWST has helped address these issues with high-quality data, it has not entirely resolved them.

Finally, a major issue this letter does not discuss is that of \emph{which models to compare}.  The assumption for Bayes factors that one of the models is correct is a very strong one, so it is important to carefully consider which models should be considered.  For example, there may be other molecules which have similar effects on the spectrum to the ones considered \citep[e.g.][]{Welbanks2025}.  This is arguably the central problem of exoplanet spectroscopy and no statistical analysis can sidestep it.

\subsection*{Acknowledgments}
Some of the data presented in this article were obtained from the Mikulski Archive for Space Telescopes (MAST) at the Space Telescope Science Institute. The specific observations analyzed can be accessed via \dataset[doi: 10.17909/3tmh-4209]{https://doi.org/10.17909/3tmh-4209}.

\appendix
\section{Python Implementation}\label{sec:python}
The following Python functions are included to assist researchers in conducting model comparisons in the ways suggested in the article.
\begin{lstlisting}[language=Python]
import numpy as np
from scipy.special import logsumexp


def get_bic(max_logl, n_data, n_params):
    '''Calculates the BIC given the max likelihood and number of data
        and parameters.  If you have a minimum chi-squared, it may be converted
        using log(L) = -chiSq/2 - sum_i(log(2*pi*sigma_i)/2), where the logs are
        in base e and sigma_i is the uncertainty for each datapoint; the latter
        term cancels out in BIC comparisons if the same data is used, so can
        usually be omitted. Citation Schwartz (1978), DOI 10.1214/aos/1176344136
    Args:
        max_logl: the maximum likelihood of the model.
        n_data: the number of datapoints the model was fit to.
        n_params: the number of parameters in the model.
    Returns:
        The BIC statistic for the model.'''
    return -2*max_logl + n_params*np.log(n_data)


def get_aic(max_logl, n_params):
    '''Calculates the AIC given the max likelihood and number of
        parameters.  If you have a minimum chi-squared, it may be converted
        using log(L) = -chiSq/2 - sum_i(log(2*pi*sigma_i)/2), where the logs are
        in base e and sigma_i is the uncertainty for each datapoint; the latter
        cancels out in AIC comparisons if the same data is used, so can usually
        be omitted.  Citation: Akaike (1974), DOI 10.1109/TAC.1974.1100705
    Args:
        max_logl: the maximum likelihood of the model.
        n_params: the number of parameters in the model.
    Returns:
        The AIC statistic for the model.'''
    return -2*max_logl + 2*n_params


def get_waic(pointwise_logl):
    '''Computes the WAIC for a model given the pointwise log-likelihood.  This
        is not normally recorded by MCMC codes so you will need to specifically
        preserve it.  In Dynesty this can be done by setting blob=True in the
        sampler initialization and modifying the likelihood function to:
            [...]
            pointwiseLogl = stats.norm.logpdf(yObserved, loc=yModel, scale=errors)
            return pointwiseLogl.sum(), pointwiseLogl
        The pointwise log likelihood can the be retrieved from the results
        object via the "blob" attribute.  EMCEE has a similar system.
        Citation: Watanabe (2010) [no DOI]
    Args:
        pointwise_logl: the likelihood for each point and posterior sample.  It
            should be a numpy array shaped like (n_points, n_samples).
    Returns:
        The WAIC statistic for the model.'''
    assert pointwise_logl.shape[0] > pointwise_logl.shape[1], \
        "I don't believe you have more points than posterior samples"
    fit_term = logsumexp(pointwise_logl, axis=0, b=1./pointwise_logl.shape[0])
    penalty_term = np.var(pointwise_logl, axis=0)
    return -2*(np.sum(fit_term) - np.sum(penalty_term))


def get_bpics(logl_samples, n_params, log_weights=None):
    '''Calculates the simplified Bayesian Predictive Information Criterion
        (BPICs) described in Ando (2011) for the given sample log-likelihoods
        and number of parameters.  Models with a lower BPICS are preferred.
        Citation: Ando(2011), DOI 10.1080/01966324.2011.10737798
    Args:
        logl_samples: the natural log of the likelihood of posterior draws
            from an MCMC run of the model.
        n_params: The number of parameters for the model.
        log_weights: the weights of the samples given, mainly for nested
            sampling posteriors.  For equally-weighted samples, leave as None.
    Returns:
        The computed BPICS as a float.'''
    weights = None
    if log_weights is not None:
        weights = np.exp(log_weights - np.max(log_weights))
        weights = (weights / np.sum(weights))
    mean_logl = np.average(logl_samples, weights=weights)
    return -2*mean_logl + 2*n_params

def get_dic(logl_samples, logl_at_mean, alternate_pd=False):
    '''Calculates the Deviance Information Criterion (DIC) for the given sample
        log-liklihoods, using the Ando (2011) variant and the Gelman (2014) number
        of effective parameters formula.  Models with lower DIC are preferred.
    Args:
        logl_samples: the natural log of the likelihood of posterior draws
            from the MCMC run; should be an array of length [nsamples].
        loglike_at_mean: the natural log of the likelihood at the at the posterior
            mean of the parameters (not the mean or max of the log likelihood!)
        alternate_pd: A flag indicating whether to use the Gelman (2014)
            effective parameters formula rather than the default Spiegelhalter
            (2012) formula.  Strictly positive but somewhat less numerically stable,
            if used ensure you have a well-converged posterior with plenty of samples.
    Returns:
        The computed DIC as a float.'''
    meanLikelihood = np.mean(logl_samples)
    if alternate_pd:
        nEffectiveParams = 2*np.var(logl_samples)
    else:
        nEffectiveParams = 2*logl_at_mean - 2*meanLikelihood
    return -2*meanLikelihood + 3*nEffectiveParams
\end{lstlisting}

\newpage
\section{Atmosphere Posteriors}
\begin{figure}[!h]
    \centering
    \includegraphics[width=\textwidth]{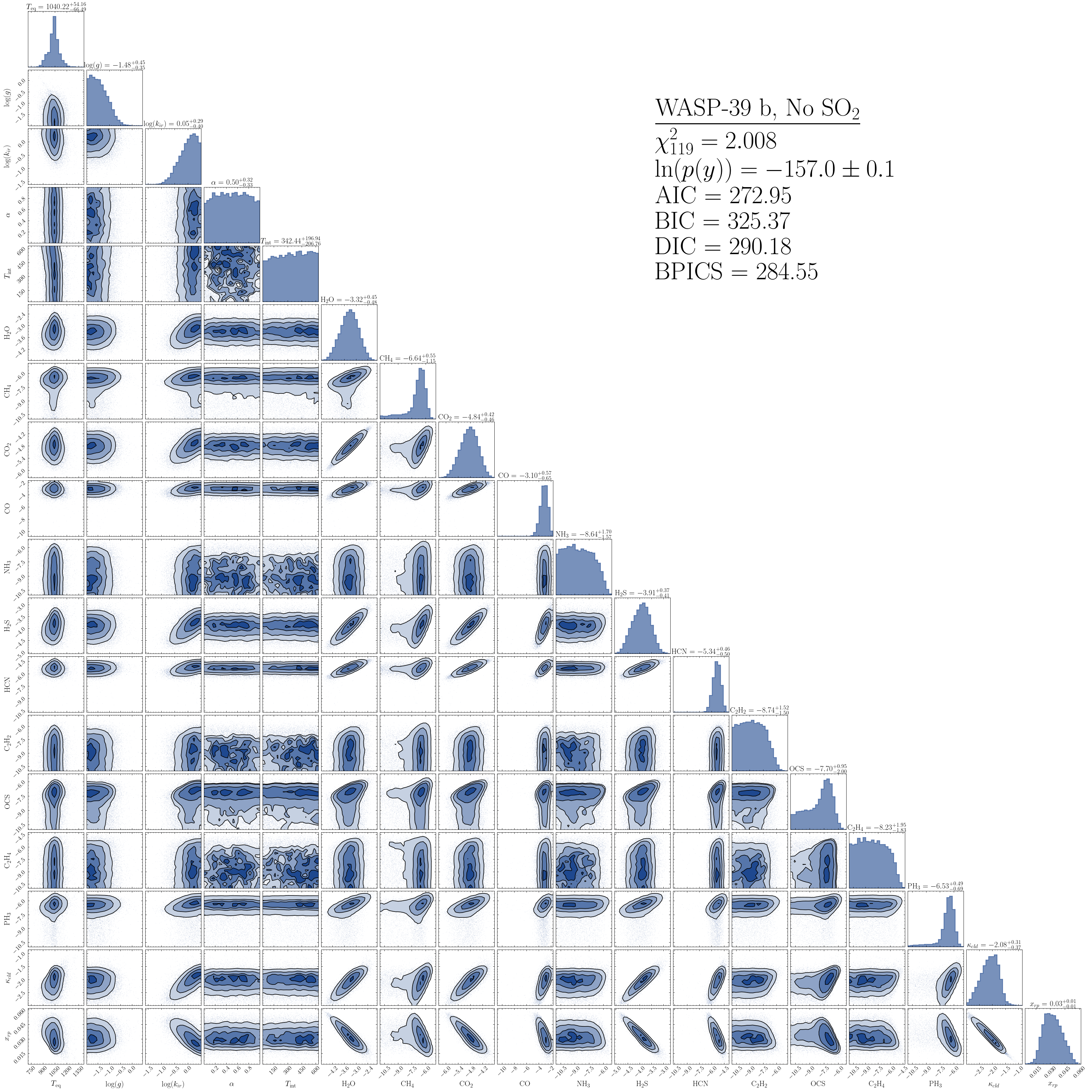}
    \caption{The posterior of the atmosphere model without SO$_2$ fitted to the  WASP-39 b transit spectrum, as described in Sec. \ref{sec:retrieval}.  Selected model comparison statistics are listed as well.}
    \label{fig:postNoSO2}
\end{figure}

\newpage
\begin{figure}[!h]
    \centering
    \includegraphics[width=\textwidth]{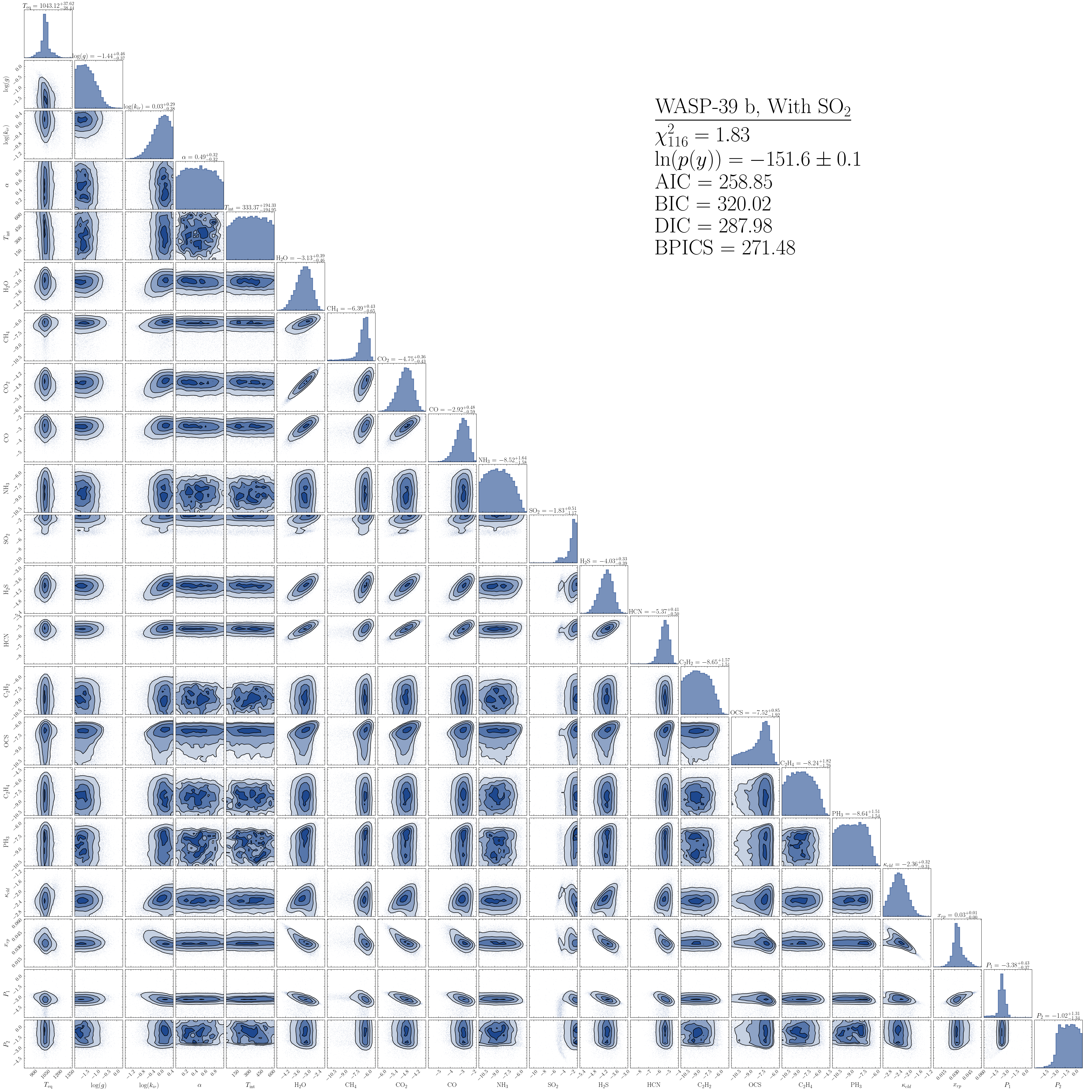}
    \caption{The posterior of the atmosphere model including SO$_2$ fitted to the  WASP-39 b transit spectrum, as described in Sec. \ref{sec:retrieval}.  Selected model comparison statistics are listed as well.}
    \label{fig:postSO2}
\end{figure}

\bibliography{bibliography}

\begin{thebibliography}{}
\expandafter\ifx\csname natexlab\endcsname\relax\def\natexlab#1{#1}\fi
\providecommand{\url}[1]{\href{#1}{#1}}
\providecommand{\dodoi}[1]{doi:~\href{http://doi.org/#1}{\nolinkurl{#1}}}
\providecommand{\doeprint}[1]{\href{http://ascl.net/#1}{\nolinkurl{http://ascl.net/#1}}}
\providecommand{\doarXiv}[1]{\href{https://arxiv.org/abs/#1}{\nolinkurl{https://arxiv.org/abs/#1}}}

\bibitem[{Adams {et~al.}(2023)Adams, Meyer, Howe, Burningham, Daemgen, Fortney, Line, Marley, Quanz, \& Todorov}]{Adams2023}
Adams, A.~D., Meyer, M.~R., Howe, A.~R., {et~al.} 2023, AJ, 166, 192, \dodoi{10.3847/1538-3881/acfb87}

\bibitem[{Ahrer {et~al.}(2023)Ahrer, Stevenson, Mansfield, Moran, Brande, Morello, Murray, Nikolov, {Petit dit de la Roche}, Schlawin, Wheatley, Zieba, Batalha, Damiano, Goyal, Lendl, Lothringer, Mukherjee, Ohno, Batalha, Battley, Bean, Beatty, Benneke, {Berta-Thompson}, Carter, Cubillos, Daylan, Espinoza, Gao, Gibson, Gill, Harrington, Hu, Kreidberg, Lewis, Line, {L{\'o}pez-Morales}, Parmentier, Powell, Sing, Tsai, Wakeford, Welbanks, Alam, Alderson, Allen, Anderson, Barstow, Bayliss, Bell, Blecic, Bryant, Burleigh, Carone, Casewell, Changeat, Chubb, Crossfield, Crouzet, Decin, D{\'e}sert, Feinstein, Flagg, Fortney, Gizis, Heng, Iro, Kempton, Kendrew, Kirk, Knutson, Komacek, Lagage, Leconte, {Lustig-Yaeger}, MacDonald, Mancini, May, Mayne, Miguel, {Mikal-Evans}, Molaverdikhani, Palle, Piaulet, Rackham, Redfield, Rogers, Roy, Rustamkulov, Shkolnik, Sotzen, Taylor, Tremblin, Tucker, Turner, {de Val-Borro}, Venot, \& Zhang}]{Ahrer2023}
Ahrer, E.-M., Stevenson, K.~B., Mansfield, M., {et~al.} 2023, Nature, 614, 653, \dodoi{10.1038/s41586-022-05590-4}

\bibitem[{Akaike(1974)}]{Akaike1974}
Akaike, H. 1974, IEEE Transactions on Automatic Control, 19, 716

\bibitem[{Akaike(1978)}]{Akaike1978}
---. 1978, Ann Inst Stat Math, 30, 9, \dodoi{10.1007/BF02480194}

\bibitem[{Akaike(1981)}]{Akaike1981}
---. 1981, Journal of Econometrics, 16, 3, \dodoi{10.1016/0304-4076(81)90071-3}

\bibitem[{Alderson {et~al.}(2022)Alderson, Wakeford, MacDonald, Lewis, May, Grant, Sing, Stevenson, Fowler, Goyal, Batalha, \& Kataria}]{Alderson2022}
Alderson, L., Wakeford, H.~R., MacDonald, R.~J., {et~al.} 2022, Monthly Notices of the Royal Astronomical Society, 512, 4185, \dodoi{10.1093/mnras/stac661}

\bibitem[{Alderson {et~al.}(2023)Alderson, Wakeford, Alam, Batalha, Lothringer, Adams~Redai, Barat, Brande, Damiano, Daylan, Espinoza, Flagg, Goyal, Grant, Hu, Inglis, Lee, {Mikal-Evans}, {Ramos-Rosado}, Roy, Wallack, Batalha, Bean, Benneke, {Berta-Thompson}, Carter, Changeat, Col{\'o}n, Crossfield, D{\'e}sert, {Foreman-Mackey}, Gibson, Kreidberg, Line, {L{\'o}pez-Morales}, Molaverdikhani, Moran, Morello, Moses, Mukherjee, Schlawin, Sing, Stevenson, Taylor, Aggarwal, Ahrer, Allen, Barstow, Bell, Blecic, Casewell, Chubb, Crouzet, Cubillos, Decin, Feinstein, Fortney, Harrington, Heng, Iro, Kempton, Kirk, Knutson, Krick, Leconte, Lendl, MacDonald, Mancini, Mansfield, May, Mayne, Miguel, Nikolov, Ohno, Palle, Parmentier, {Petit dit de la Roche}, Piaulet, Powell, Rackham, Redfield, Rogers, Rustamkulov, Tan, Tremblin, Tsai, Turner, {de Val-Borro}, Venot, Welbanks, Wheatley, \& Zhang}]{Alderson2023}
Alderson, L., Wakeford, H.~R., Alam, M.~K., {et~al.} 2023, Nature, 614, 664, \dodoi{10.1038/s41586-022-05591-3}

\bibitem[{Anderson {et~al.}(1998)Anderson, Burnham, \& White}]{Anderson1998}
Anderson, D.~R., Burnham, K.~P., \& White, G.~C. 1998, Journal of Applied Statistics, 25, 263, \dodoi{10.1080/02664769823250}

\bibitem[{Ando(2007)}]{Ando2007}
Ando, T. 2007, Biometrika, 94, 443

\bibitem[{Ando(2011)}]{Ando2011}
---. 2011, American Journal of Mathematical and Management Sciences, 31, 13, \dodoi{10.1080/01966324.2011.10737798}

\bibitem[{Batalha {et~al.}(2019)Batalha, Marley, Lewis, \& Fortney}]{Batalha2019}
Batalha, N.~E., Marley, M.~S., Lewis, N.~K., \& Fortney, J.~J. 2019, ApJ, 878, 70, \dodoi{10.3847/1538-4357/ab1b51}

\bibitem[{Bayes \& Price(1763)}]{Bayes1763}
Bayes, T., \& Price, R. 1763, Philosophical Transactions of the Royal Society of London Series I, 53, 370

\bibitem[{Bell {et~al.}(2023)Bell, Welbanks, Schlawin, Line, Fortney, Greene, Ohno, Parmentier, Rauscher, Beatty, Mukherjee, Wiser, Boyer, Rieke, \& Stansberry}]{Bell2023}
Bell, T.~J., Welbanks, L., Schlawin, E., {et~al.} 2023, Nature, 623, 709, \dodoi{10.1038/s41586-023-06687-0}

\bibitem[{Benneke \& Seager(2012)}]{Benneke2012}
Benneke, B., \& Seager, S. 2012, The Astrophysical Journal, 753, 100, \dodoi{10.1088/0004-637X/753/2/100}

\bibitem[{Benneke \& Seager(2013)}]{Benneke2013}
---. 2013, The Astrophysical Journal, 778, 153, \dodoi{10.1088/0004-637X/778/2/153}

\bibitem[{Benneke {et~al.}(2019)Benneke, Wong, Piaulet, Knutson, Lothringer, Morley, Crossfield, Gao, Greene, Dressing, Dragomir, Howard, McCullough, Kempton, Fortney, \& Fraine}]{Benneke2019}
Benneke, B., Wong, I., Piaulet, C., {et~al.} 2019, The Astrophysical Journal, 887, L14, \dodoi{10.3847/2041-8213/ab59dc}

\bibitem[{Bevington \& Robinson(1992)}]{Bevington1992}
Bevington, P.~R., \& Robinson, D.~K. 1992, Data Reduction and Error Analysis for the Physical Sciences (New York: McGraw-Hill)

\bibitem[{Box(1980)}]{Box1980}
Box, G. E.~P. 1980, Journal of the Royal Statistical Society. Series A (General), 143, 383, \dodoi{10.2307/2982063}

\bibitem[{Bozdogan(1987)}]{Bozdogan1987}
Bozdogan, H. 1987, Psychometrika, 52, 345, \dodoi{10.1007/BF02294361}

\bibitem[{Brande {et~al.}(2022)Brande, Crossfield, Kreidberg, Oklop{\v c}i{\'c}, Polanski, Barman, Benneke, Christiansen, Dragomir, {Foreman-Mackey}, Fortney, Greene, Howard, Knutson, Lothringer, {Mikal-Evans}, \& Morley}]{Brande2022}
Brande, J., Crossfield, I. J.~M., Kreidberg, L., {et~al.} 2022, The Astronomical Journal, 164, 197, \dodoi{10.3847/1538-3881/ac8b7e}

\bibitem[{Buchner {et~al.}(2014)Buchner, Georgakakis, Nandra, Hsu, Rangel, Brightman, Merloni, Salvato, Donley, \& Kocevski}]{Buchner2014}
Buchner, J., Georgakakis, A., Nandra, K., {et~al.} 2014, A\&A, 564, A125, \dodoi{10.1051/0004-6361/201322971}

\bibitem[{Burnham {et~al.}(2011)Burnham, Anderson, \& Huyvaert}]{Burnham2011}
Burnham, K.~P., Anderson, D.~R., \& Huyvaert, K.~P. 2011, Behav Ecol Sociobiol, 65, 23, \dodoi{10.1007/s00265-010-1029-6}

\bibitem[{Carone {et~al.}(2021)Carone, Molli{\`e}re, Zhou, Bouwman, Yan, Baeyens, Apai, Espinoza, Rackham, Jord{\'a}n, Angerhausen, Decin, Lendl, Venot, \& Henning}]{Carone2021}
Carone, L., Molli{\`e}re, P., Zhou, Y., {et~al.} 2021, A\&A, 646, A168, \dodoi{10.1051/0004-6361/202038620}

\bibitem[{Chen {et~al.}(2021)Chen, Pall{\'e}, Parviainen, Murgas, \& Yan}]{Chen2021}
Chen, G., Pall{\'e}, E., Parviainen, H., Murgas, F., \& Yan, F. 2021, The Astrophysical Journal, 913, L16, \dodoi{10.3847/2041-8213/abfbe1}

\bibitem[{Chen \& Rogers(2016)}]{Chen2016}
Chen, H., \& Rogers, L.~A. 2016, The Astrophysical Journal, 831, 180, \dodoi{10.3847/0004-637X/831/2/180}

\bibitem[{Damiano {et~al.}(2024)Damiano, {Bello-Arufe}, Yang, \& Hu}]{Damiano2024}
Damiano, M., {Bello-Arufe}, A., Yang, J., \& Hu, R. 2024, The Astrophysical Journal, 968, L22, \dodoi{10.3847/2041-8213/ad5204}

\bibitem[{De~Regt {et~al.}(2024)De~Regt, Gandhi, Snellen, Zhang, Ginski, Gonz{\'a}lez~Picos, Kesseli, Landman, Molli{\`e}re, Nasedkin, {S{\'a}nchez-L{\'o}pez}, \& Stolker}]{DeRegt2024}
De~Regt, S., Gandhi, S., Snellen, I. A.~G., {et~al.} 2024, A\&A, 688, A116, \dodoi{10.1051/0004-6361/202348508}

\bibitem[{Del~Genio {et~al.}(2019)Del~Genio, Kiang, Way, Amundsen, Sohl, Fujii, Chandler, Aleinov, Colose, Guzewich, \& Kelley}]{DelGenio2019}
Del~Genio, A.~D., Kiang, N.~Y., Way, M.~J., {et~al.} 2019, ApJ, 884, 75, \dodoi{10.3847/1538-4357/ab3be8}

\bibitem[{Feng {et~al.}(2016)Feng, Line, Fortney, Stevenson, Bean, Kreidberg, \& Parmentier}]{Feng2016}
Feng, Y.~K., Line, M.~R., Fortney, J.~J., {et~al.} 2016, The Astrophysical Journal, 829, 52, \dodoi{10.3847/0004-637X/829/1/52}

\bibitem[{Fisher \& Heng(2018)}]{Fisher2018}
Fisher, C., \& Heng, K. 2018, Monthly Notices of the Royal Astronomical Society, 481, 4698, \dodoi{10.1093/mnras/sty2550}

\bibitem[{Ford \& Gregory(2006)}]{Ford2006a}
Ford, E.~B., \& Gregory, P.~C. 2006, Bayesian {{Model Selection}} and {{Extrasolar Planet Detection}},  arXiv, \dodoi{10.48550/arXiv.astro-ph/0608328}

\bibitem[{{Foreman-Mackey} {et~al.}(2013){Foreman-Mackey}, Hogg, Lang, \& Goodman}]{Foreman-Mackey2013}
{Foreman-Mackey}, D., Hogg, D.~W., Lang, D., \& Goodman, J. 2013, Publications of the Astronomical Society of the Pacific, 125, 306, \dodoi{10.1086/670067}

\bibitem[{Fraine {et~al.}(2014)Fraine, Deming, Benneke, Knutson, Jord{\'a}n, Espinoza, Madhusudhan, Wilkins, \& Todorov}]{Fraine2014}
Fraine, J., Deming, D., Benneke, B., {et~al.} 2014, Nature, 513, 526, \dodoi{10.1038/nature13785}

\bibitem[{Fu {et~al.}(2024)Fu, Welbanks, Deming, Inglis, Zhang, Lothringer, Ih, Moses, Schlawin, Knutson, Henry, Greene, Sing, Savel, Kempton, Louie, Line, \& Nixon}]{Fu2024}
Fu, G., Welbanks, L., Deming, D., {et~al.} 2024, Nature, 632, 752, \dodoi{10.1038/s41586-024-07760-y}

\bibitem[{Gelman {et~al.}(2014)Gelman, Carlin, Stern, Dunson, Vehtari, \& Rubin}]{Gelman2014}
Gelman, A., Carlin, J.~B., Stern, H.~S., {et~al.} 2014, Bayesian Data Analysis, third edition edn., Texts in Statistical Science Series (Boca Raton London New York: {CRC Press, Taylor and Francis Group})

\bibitem[{Gelman \& Rubin(1992)}]{Gelman1992}
Gelman, A., \& Rubin, D.~B. 1992, Statist. Sci., 7, 457, \dodoi{10.1214/ss/1177011136}

\bibitem[{Gibson {et~al.}(2010)Gibson, Aigrain, Pollacco, Barros, Hebb, Hrudkov{\'a}, Simpson, Skillen, \& West}]{Gibson2010}
Gibson, N.~P., Aigrain, S., Pollacco, D.~L., {et~al.} 2010, Monthly Notices of the Royal Astronomical Society, 404, L114, \dodoi{10.1111/j.1745-3933.2010.00847.x}

\bibitem[{Gibson {et~al.}(2012)Gibson, Aigrain, Pont, Sing, D{\'e}sert, Evans, Henry, Husnoo, \& Knutson}]{Gibson2012}
Gibson, N.~P., Aigrain, S., Pont, F., {et~al.} 2012, Monthly Notices of the Royal Astronomical Society, 422, 753, \dodoi{10.1111/j.1365-2966.2012.20655.x}

\bibitem[{Grasser {et~al.}(2025)Grasser, Snellen, De~Regt, Gonz{\'a}lez~Picos, Zhang, Stolker, Gandhi, Nasedkin, Landman, Kesseli, \& Mulder}]{Grasser2025}
Grasser, N., Snellen, I. A.~G., De~Regt, S., {et~al.} 2025, A\&A, 698, A252, \dodoi{10.1051/0004-6361/202554195}

\bibitem[{Gressier {et~al.}(2024)Gressier, Espinoza, Allen, Sing, Banerjee, Barstow, Valenti, Lewis, Birkmann, Challener, Manjavacas, Alves De~Oliveira, Crouzet, \& Beck}]{Gressier2024}
Gressier, A., Espinoza, N., Allen, N.~H., {et~al.} 2024, ApJL, 975, L10, \dodoi{10.3847/2041-8213/ad73d1}

\bibitem[{Guillot(2010)}]{Guillot2010}
Guillot, T. 2010, Astronomy and Astrophysics, 520, A27, \dodoi{10.1051/0004-6361/200913396}

\bibitem[{Guilluy {et~al.}(2021)Guilluy, Gressier, Wright, Santerne, Jaziri, Edwards, Changeat, {Modirrousta-Galian}, Skaf, {Al-Refaie}, Baeyens, Bieger, Blain, Kiefer, Morvan, Mugnai, Pluriel, Poveda, Zingales, Whiteford, Yip, Charnay, Leconte, Drossart, Sozzetti, Marcq, Tsiaras, Venot, Waldmann, \& Beaulieu}]{Guilluy2021}
Guilluy, G., Gressier, A., Wright, S., {et~al.} 2021, The Astronomical Journal, 161, 19, \dodoi{10.3847/1538-3881/abc3c8}

\bibitem[{Hoeijmakers {et~al.}(2018)Hoeijmakers, Ehrenreich, Heng, Kitzmann, Grimm, Allart, Deitrick, Wyttenbach, Oreshenko, Pino, Rimmer, Molinari, \& Di~Fabrizio}]{Hoeijmakers2018}
Hoeijmakers, H.~J., Ehrenreich, D., Heng, K., {et~al.} 2018, arXiv:1808.05653 [astro-ph, physics:physics].
\newblock \doarXiv{1808.05653}

\bibitem[{Jeffreys(1935)}]{Jeffreys1935}
Jeffreys, H. 1935, Mathematical Proceedings of the Cambridge Philosophical Society, 31, 203, \dodoi{10.1017/S030500410001330X}

\bibitem[{Jeffreys(1939)}]{Jeffreys1939}
---. 1939, Theory of Probability (Oxford, England: Clarendon Press)

\bibitem[{Jord{\'a}n {et~al.}(2013)Jord{\'a}n, Espinoza, Rabus, Eyheramendy, Sing, D{\'e}sert, Bakos, Fortney, {L{\'o}pez-Morales}, Maxted, Triaud, \& Szentgyorgyi}]{Jordan2013}
Jord{\'a}n, A., Espinoza, N., Rabus, M., {et~al.} 2013, ApJ, 778, 184, \dodoi{10.1088/0004-637X/778/2/184}

\bibitem[{{JWST Transiting Exoplanet Community Early Release Science Team} {et~al.}(2023){JWST Transiting Exoplanet Community Early Release Science Team}, Ahrer, Alderson, Batalha, Batalha, Bean, Beatty, Bell, Benneke, {Berta-Thompson}, Carter, Crossfield, Espinoza, Feinstein, Fortney, Gibson, Goyal, Kempton, Kirk, Kreidberg, {L{\'o}pez-Morales}, Line, Lothringer, Moran, Mukherjee, Ohno, Parmentier, Piaulet, Rustamkulov, Schlawin, Sing, Stevenson, Wakeford, Allen, Birkmann, Brande, Crouzet, Cubillos, Damiano, D{\'e}sert, Gao, Harrington, Hu, Kendrew, Knutson, Lagage, Leconte, Lendl, MacDonald, May, Miguel, Molaverdikhani, Moses, Murray, Nehring, Nikolov, {Petit dit de la Roche}, Radica, Roy, Stassun, Taylor, Waalkes, Wachiraphan, Welbanks, Wheatley, Aggarwal, Alam, Banerjee, Barstow, Blecic, Casewell, Changeat, Chubb, Col{\'o}n, Coulombe, Daylan, {de Val-Borro}, Decin, Dos~Santos, Flagg, France, Fu, Garc{\'i}a~Mu{\~n}oz, Gizis, Glidden, Grant, Heng, Henning, Hong, Inglis, Iro, Kataria, Komacek, Krick, Lee,
  Lewis, {Lillo-Box}, {Lustig-Yaeger}, Mancini, Mandell, Mansfield, Marley, {Mikal-Evans}, Morello, Nixon, Ortiz~Ceballos, Piette, Powell, Rackham, {Ramos-Rosado}, Rauscher, Redfield, Rogers, Roman, Roudier, Scarsdale, Shkolnik, Southworth, Spake, Steinrueck, Tan, Teske, Tremblin, Tsai, Tucker, Turner, Valenti, Venot, Waldmann, Wallack, Zhang, \& Zieba}]{jwstCollab2023}
{JWST Transiting Exoplanet Community Early Release Science Team}, Ahrer, E.-M., Alderson, L., {et~al.} 2023, Nature, 614, 649, \dodoi{10.1038/s41586-022-05269-w}

\bibitem[{Kang {et~al.}(2024)Kang, Chen, Jiang, Pall{\'e}, Murgas, Parviainen, Ma, Fukui, \& Narita}]{Kang2024}
Kang, H., Chen, G., Jiang, C., {et~al.} 2024, A\&A, 687, A9, \dodoi{10.1051/0004-6361/202449915}

\bibitem[{Kass \& Raftery(1995)}]{Kass1995}
Kass, R.~E., \& Raftery, A.~E. 1995, Journal of the American Statistical Association, 90, 773, \dodoi{10.1080/01621459.1995.10476572}

\bibitem[{Kass \& Wasserman(1995)}]{Kass1995a}
Kass, R.~E., \& Wasserman, L. 1995, Journal of the American Statistical Association, 90, 928, \dodoi{10.1080/01621459.1995.10476592}

\bibitem[{Kempton {et~al.}(2023)Kempton, Zhang, Bean, Steinrueck, Piette, Parmentier, Malsky, Roman, Rauscher, Gao, Bell, Xue, Taylor, Savel, Arnold, Nixon, Stevenson, Mansfield, Kendrew, Zieba, Ducrot, Dyrek, Lagage, Stassun, Henry, Barman, Lupu, Malik, Kataria, Ih, Fu, Welbanks, \& McGill}]{Kempton2023a}
Kempton, E. M.~R., Zhang, M., Bean, J.~L., {et~al.} 2023, Nature, 620, 67, \dodoi{10.1038/s41586-023-06159-5}

\bibitem[{Kipping \& Benneke(2025)}]{Kipping2025}
Kipping, D., \& Benneke, B. 2025, Exoplaneteers {{Keep Overestimating Sigma Significances}},  arXiv, \dodoi{10.48550/arXiv.2506.05392}

\bibitem[{Kipping {et~al.}(2013)Kipping, Forgan, Hartman, Nesvorn{\'y}, Bakos, Schmitt, \& Buchhave}]{Kipping2013}
Kipping, D.~M., Forgan, D., Hartman, J., {et~al.} 2013, ApJ, 777, 134, \dodoi{10.1088/0004-637X/777/2/134}

\bibitem[{Konishi \& Kitagawa(2008)}]{Konishi2008}
Konishi, S., \& Kitagawa, G. 2008, Information {{Criteria}} and {{Statistical Modeling}}, Springer {{Series}} in {{Statistics}} (New York, NY: Springer New York), \dodoi{10.1007/978-0-387-71887-3}

\bibitem[{Koposov {et~al.}(2024)Koposov, Speagle, Barbary, Ashton, Bennett, Buchner, Scheffler, Cook, Talbot, Guillochon, Cubillos, Ramos, Dartiailh, {Ilya}, Tollerud, Lang, Johnson, {jtmendel}, Higson, Vandal, Daylan, Angus, {patelR}, Cargile, Sheehan, Pitkin, Kirk, Leja, {joezuntz}, \& Goldstein}]{Koposov2024}
Koposov, S., Speagle, J., Barbary, K., {et~al.} 2024, Joshspeagle/Dynesty: V2.1.4, Zenodo, \dodoi{10.5281/zenodo.12537467}

\bibitem[{Kreidberg {et~al.}(2014)Kreidberg, Bean, D{\'e}sert, Benneke, Deming, Stevenson, Seager, {Berta-Thompson}, Seifahrt, \& Homeier}]{Kreidberg2014}
Kreidberg, L., Bean, J.~L., D{\'e}sert, J.-M., {et~al.} 2014, Nature, 505, 69, \dodoi{10.1038/nature12888}

\bibitem[{Kreidberg {et~al.}(2015)Kreidberg, Line, Bean, Stevenson, D{\'e}sert, Madhusudhan, Fortney, Barstow, Henry, Williamson, \& Showman}]{Kreidberg2015a}
Kreidberg, L., Line, M.~R., Bean, J.~L., {et~al.} 2015, The Astrophysical Journal, 814, 66, \dodoi{10.1088/0004-637X/814/1/66}

\bibitem[{K{\"u}hnle {et~al.}(2025)K{\"u}hnle, Patapis, Molli{\`e}re, Tremblin, Matthews, Glauser, Whiteford, Vasist, Absil, Barrado, Min, Lagage, Waters, Guedel, Henning, Vandenbussche, Baudoz, Decin, Pye, Royer, Van~Dishoeck, {\"O}stlin, Ray, \& Wright}]{Kuhnle2025}
K{\"u}hnle, H., Patapis, P., Molli{\`e}re, P., {et~al.} 2025, A\&A, 695, A224, \dodoi{10.1051/0004-6361/202452547}

\bibitem[{Lehmann \& Romano(2022)}]{Lehmann2022}
Lehmann, E.~L., \& Romano, J.~P. 2022, Testing Statistical Hypotheses, forth edition edn., Springer {{Texts}} in {{Statistics}} (Cham: Springer), \dodoi{10.1007/978-3-030-70578-7}

\bibitem[{Line {et~al.}(2016)Line, Stevenson, Bean, Desert, Fortney, Kreidberg, Madhusudhan, Showman, \& {Diamond-Lowe}}]{Line2016a}
Line, M.~R., Stevenson, K.~B., Bean, J., {et~al.} 2016, The Astronomical Journal, 152, 203, \dodoi{10.3847/0004-6256/152/6/203}

\bibitem[{Loredo {et~al.}(2012)Loredo, Berger, Chernoff, Clyde, \& Liu}]{Loredo2012}
Loredo, T.~J., Berger, J.~O., Chernoff, D.~F., Clyde, M.~A., \& Liu, B. 2012, Statistical Methodology, 9, 101, \dodoi{10.1016/j.stamet.2011.07.005}

\bibitem[{MacDonald \& Madhusudhan(2017)}]{MacDonald2017}
MacDonald, R.~J., \& Madhusudhan, N. 2017, Monthly Notices of the Royal Astronomical Society, 469, 1979, \dodoi{10.1093/mnras/stx804}

\bibitem[{Madhusudhan {et~al.}(2025)Madhusudhan, Constantinou, Holmberg, Sarkar, Piette, \& Moses}]{Madhusudhan2025}
Madhusudhan, N., Constantinou, S., Holmberg, M., {et~al.} 2025, ApJL, 983, L40, \dodoi{10.3847/2041-8213/adc1c8}

\bibitem[{Madhusudhan {et~al.}(2020)Madhusudhan, Nixon, Welbanks, Piette, \& Booth}]{Madhusudhan2020}
Madhusudhan, N., Nixon, M.~C., Welbanks, L., Piette, A. A.~A., \& Booth, R.~A. 2020, The Astrophysical Journal, 891, L7, \dodoi{10.3847/2041-8213/ab7229}

\bibitem[{Madhusudhan {et~al.}(2011)Madhusudhan, Harrington, Stevenson, Nymeyer, Campo, Wheatley, Deming, Blecic, Hardy, Lust, Anderson, {Collier-Cameron}, Britt, Bowman, Hebb, Hellier, Maxted, Pollacco, \& West}]{Madhusudhan2011a}
Madhusudhan, N., Harrington, J., Stevenson, K.~B., {et~al.} 2011, Nature, 469, 64, \dodoi{10.1038/nature09602}

\bibitem[{Mancini {et~al.}(2018)Mancini, Esposito, Covino, Southworth, Biazzo, Bruni, Ciceri, Evans, Lanza, Poretti, Sarkis, Smith, Brogi, Affer, Benatti, Bignamini, Boccato, Bonomo, Borsa, Carleo, Claudi, Cosentino, Damasso, Desidera, Giacobbe, {Gonz{\'a}lez-{\'A}lvarez}, Gratton, Harutyunyan, Leto, Maggio, Malavolta, Maldonado, {Martinez-Fiorenzano}, Masiero, Micela, Molinari, Nascimbeni, Pagano, Pedani, Piotto, Rainer, Scandariato, Smareglia, Sozzetti, Andreuzzi, \& Henning}]{Mancini2018}
Mancini, L., Esposito, M., Covino, E., {et~al.} 2018, A\&A, 613, A41, \dodoi{10.1051/0004-6361/201732234}

\bibitem[{Mandell {et~al.}(2013)Mandell, Haynes, Sinukoff, Madhusudhan, Burrows, \& Deming}]{Mandell2013}
Mandell, A.~M., Haynes, K., Sinukoff, E., {et~al.} 2013, The Astrophysical Journal, 779, 128, \dodoi{10.1088/0004-637X/779/2/128}

\bibitem[{May {et~al.}(2023)May, MacDonald, Bennett, Moran, Wakeford, Peacock, {Lustig-Yaeger}, Highland, Stevenson, Sing, Mayorga, Batalha, Kirk, {L{\'o}pez-Morales}, Valenti, Alam, Alderson, Fu, {Gonzalez-Quiles}, Lothringer, Rustamkulov, \& Sotzen}]{May2023}
May, E.~M., MacDonald, R.~J., Bennett, K.~A., {et~al.} 2023, The Astrophysical Journal, 959, L9, \dodoi{10.3847/2041-8213/ad054f}

\bibitem[{McCreery {et~al.}(2025)McCreery, Dos~Santos, Espinoza, Allart, \& Kirk}]{McCreery2025}
McCreery, P., Dos~Santos, L.~A., Espinoza, N., Allart, R., \& Kirk, J. 2025, ApJ, 980, 125, \dodoi{10.3847/1538-4357/ada6b9}

\bibitem[{Meng(1994)}]{Meng1994}
Meng, X.-L. 1994, Ann. Statist., 22, \dodoi{10.1214/aos/1176325622}

\bibitem[{{Mikal-Evans} {et~al.}(2021){Mikal-Evans}, Crossfield, Benneke, Kreidberg, Moses, Morley, Thorngren, Molli{\`e}re, {Hardegree-Ullman}, Brewer, Christiansen, Ciardi, Dragomir, Dressing, Fortney, Gorjian, Greene, Hirsch, Howard, Howell, Isaacson, Kosiarek, Krick, Livingston, Lothringer, Morales, Petigura, Schlieder, \& Werner}]{Mikal-Evans2021}
{Mikal-Evans}, T., Crossfield, I. J.~M., Benneke, B., {et~al.} 2021, The Astronomical Journal, 161, 18, \dodoi{10.3847/1538-3881/abc874}

\bibitem[{Moran {et~al.}(2018)Moran, H{\"o}rst, Batalha, Lewis, \& Wakeford}]{Moran2018}
Moran, S.~E., H{\"o}rst, S.~M., Batalha, N.~E., Lewis, N.~K., \& Wakeford, H.~R. 2018, The Astronomical Journal, 156, 252, \dodoi{10.3847/1538-3881/aae83a}

\bibitem[{Moran {et~al.}(2023)Moran, Stevenson, Sing, MacDonald, Kirk, {Lustig-Yaeger}, Peacock, Mayorga, Bennett, {L{\'o}pez-Morales}, May, Rustamkulov, Valenti, Adams~Redai, Alam, Batalha, Fu, {Gonzalez-Quiles}, Highland, Kruse, Lothringer, Ortiz~Ceballos, Sotzen, \& Wakeford}]{Moran2023}
Moran, S.~E., Stevenson, K.~B., Sing, D.~K., {et~al.} 2023, The Astrophysical Journal, 948, L11, \dodoi{10.3847/2041-8213/accb9c}

\bibitem[{Morello {et~al.}(2019)Morello, Danielski, Dickens, Tremblin, \& Lagage}]{Morello2019}
Morello, G., Danielski, C., Dickens, D., Tremblin, P., \& Lagage, P.-O. 2019, AJ, 157, 205, \dodoi{10.3847/1538-3881/ab14e2}

\bibitem[{Mukherjee {et~al.}(2023)Mukherjee, Batalha, Fortney, \& Marley}]{Mukherjee2023}
Mukherjee, S., Batalha, N.~E., Fortney, J.~J., \& Marley, M.~S. 2023, ApJ, 942, 71, \dodoi{10.3847/1538-4357/ac9f48}

\bibitem[{Mukherjee {et~al.}(2024{\natexlab{a}})Mukherjee, Fortney, Wogan, Sing, \& Ohno}]{Mukherjee2024}
Mukherjee, S., Fortney, J.~J., Wogan, N.~F., Sing, D.~K., \& Ohno, K. 2024{\natexlab{a}}, Effects of {{Planetary Parameters}} on {{Disequilibrium Chemistry}} in {{Irradiated Planetary Atmospheres}}: {{From Gas Giants}} to {{Sub-Neptunes}},  arXiv, \dodoi{10.48550/ARXIV.2410.17169}

\bibitem[{Mukherjee {et~al.}(2024{\natexlab{b}})Mukherjee, Fortney, Morley, Batalha, Marley, Karalidi, Visscher, Lupu, Freedman, \& {Gharib-Nezhad}}]{Mukherjee2024a}
Mukherjee, S., Fortney, J.~J., Morley, C.~V., {et~al.} 2024{\natexlab{b}}, ApJ, 963, 73, \dodoi{10.3847/1538-4357/ad18c2}

\bibitem[{Mulder {et~al.}(2025)Mulder, De~Regt, Landman, Picos, Snellen, Zhang, Gandhi, Ginski, Kesseli, Nasedkin, \& Stolker}]{Mulder2025}
Mulder, W., De~Regt, S., Landman, R., {et~al.} 2025, A\&A, 694, A164, \dodoi{10.1051/0004-6361/202452859}

\bibitem[{Murphy {et~al.}(2024)Murphy, Beatty, Schlawin, Bell, Line, Greene, Parmentier, Rauscher, Welbanks, Fortney, \& Rieke}]{Murphy2024}
Murphy, M.~M., Beatty, T.~G., Schlawin, E., {et~al.} 2024, Nature Astronomy, 8, 1562, \dodoi{10.1038/s41550-024-02367-9}

\bibitem[{Nasedkin {et~al.}(2024)Nasedkin, Molli{\`e}re, Lacour, Nowak, Kreidberg, Stolker, Wang, Balmer, Kammerer, Shangguan, Abuter, Amorim, {Asensio-Torres}, Benisty, Berger, Beust, Blunt, Boccaletti, Bonnefoy, Bonnet, Bordoni, Bourdarot, Brandner, Cantalloube, Caselli, Charnay, Chauvin, Chavez, Choquet, Christiaens, Cl{\'e}net, Coud{\'e} Du~Foresto, Cridland, Davies, Dembet, Dexter, Drescher, Duvert, Eckart, Eisenhauer, F{\"o}rster~Schreiber, Garcia, Garcia~Lopez, Gendron, Genzel, Gillessen, Girard, Grant, Haubois, Hei{\ss}el, Henning, Hinkley, Hippler, Houll{\'e}, Hubert, Jocou, Keppler, Kervella, Kurtovic, Lagrange, Lapeyr{\`e}re, Le~Bouquin, Lutz, Maire, Mang, Marleau, M{\'e}rand, Monnier, Mordasini, Ott, Otten, Paladini, Paumard, Perraut, Perrin, Pfuhl, Pourr{\'e}, Pueyo, Ribeiro, Rickman, Ruffio, Rustamkulov, Shimizu, Sing, Stadler, Straub, Straubmeier, Sturm, Tacconi, Van~Dishoeck, Vigan, Vincent, Von~Fellenberg, Widmann, Winterhalder, Woillez, Yazici, \& {the GRAVITY Collaboration}}]{Nasedkin2024}
Nasedkin, E., Molli{\`e}re, P., Lacour, S., {et~al.} 2024, A\&A, 687, A298, \dodoi{10.1051/0004-6361/202449328}

\bibitem[{Nelson {et~al.}(2020)Nelson, Ford, Buchner, Cloutier, D{\'i}az, Faria, Hara, Rajpaul, \& Rukdee}]{Nelson2020}
Nelson, B.~E., Ford, E.~B., Buchner, J., {et~al.} 2020, AJ, 159, 73, \dodoi{10.3847/1538-3881/ab5190}

\bibitem[{Nesseris \& {Garc{\'i}a-Bellido}(2013)}]{Nesseris2013}
Nesseris, S., \& {Garc{\'i}a-Bellido}, J. 2013, Journal of Cosmology and Astroparticle Physics, 2013, 036, \dodoi{10.1088/1475-7516/2013/08/036}

\bibitem[{{Piaulet-Ghorayeb} {et~al.}(2024){Piaulet-Ghorayeb}, Benneke, Radica, Raul, Coulombe, Ahrer, Kubyshkina, Howard, {Krissansen-Totton}, MacDonald, Roy, Louca, Christie, {Fournier-Tondreau}, Allart, Miguel, Schlichting, Welbanks, Cadieux, Dorn, {Evans-Soma}, Fortney, Pierrehumbert, Lafreniere, Acuna, Komacek, Innes, Beatty, Cloutier, Doyon, Gagnebin, Gapp, \& Knutson}]{Piaulet-Ghorayeb2024}
{Piaulet-Ghorayeb}, C., Benneke, B., Radica, M., {et~al.} 2024, ApJL, 974, L10, \dodoi{10.3847/2041-8213/ad6f00}

\bibitem[{Pont {et~al.}(2006)Pont, Zucker, \& Queloz}]{Pont2006}
Pont, F., Zucker, S., \& Queloz, D. 2006, Monthly Notices of the Royal Astronomical Society, 373, 231, \dodoi{10.1111/j.1365-2966.2006.11012.x}

\bibitem[{Powell {et~al.}(2024)Powell, Feinstein, Lee, Zhang, Tsai, Taylor, Kirk, Bell, Barstow, Gao, Bean, Blecic, Chubb, Crossfield, Jordan, Kitzmann, Moran, Morello, Moses, Welbanks, Yang, Zhang, Ahrer, {Bello-Arufe}, Brande, Casewell, Crouzet, Cubillos, Demory, Dyrek, Flagg, Hu, Inglis, Jones, Kreidberg, {L{\'o}pez-Morales}, Lagage, Meier~Vald{\'e}s, Miguel, Parmentier, Piette, Rackham, Radica, Redfield, Stevenson, Wakeford, Aggarwal, Alam, Batalha, Batalha, Benneke, {Berta-Thompson}, Brady, Caceres, Carter, D{\'e}sert, Harrington, Iro, Line, Lothringer, MacDonald, Mancini, Molaverdikhani, Mukherjee, Nixon, Oza, Palle, Rustamkulov, Sing, Steinrueck, Venot, Wheatley, \& Yurchenko}]{Powell2024}
Powell, D., Feinstein, A.~D., Lee, E. K.~H., {et~al.} 2024, Nature, 626, 979, \dodoi{10.1038/s41586-024-07040-9}

\bibitem[{Rotman {et~al.}(2025)Rotman, Welbanks, Line, McGill, Radica, \& Nixon}]{Rotman2025}
Rotman, Y., Welbanks, L., Line, M.~R., {et~al.} 2025, ApJ, 989, 201, \dodoi{10.3847/1538-4357/adef04}

\bibitem[{Rustamkulov {et~al.}(2023)Rustamkulov, Sing, Mukherjee, May, Kirk, Schlawin, Line, Piaulet, Carter, Batalha, Goyal, {L{\'o}pez-Morales}, Lothringer, MacDonald, Moran, Stevenson, Wakeford, Espinoza, Bean, Batalha, Benneke, {Berta-Thompson}, Crossfield, Gao, Kreidberg, Powell, Cubillos, Gibson, Leconte, Molaverdikhani, Nikolov, Parmentier, Roy, Taylor, Turner, Wheatley, Aggarwal, Ahrer, Alam, Alderson, Allen, Banerjee, Barat, Barrado, Barstow, Bell, Blecic, Brande, Casewell, Changeat, Chubb, Crouzet, Daylan, Decin, D{\'e}sert, {Mikal-Evans}, Feinstein, Flagg, Fortney, Harrington, Heng, Hong, Hu, Iro, Kataria, Kempton, Krick, Lendl, {Lillo-Box}, Louca, {Lustig-Yaeger}, Mancini, Mansfield, Mayne, Miguel, Morello, Ohno, Palle, Petit Dit De La~Roche, Rackham, Radica, {Ramos-Rosado}, Redfield, Rogers, Shkolnik, Southworth, Teske, Tremblin, Tucker, Venot, Waalkes, Welbanks, Zhang, \& Zieba}]{Rustamkulov2023}
Rustamkulov, Z., Sing, D.~K., Mukherjee, S., {et~al.} 2023, Nature, 614, 659, \dodoi{10.1038/s41586-022-05677-y}

\bibitem[{Schwarz(1978)}]{Schwarz1978}
Schwarz, G. 1978, Ann. Statist., 6, 461, \dodoi{10.1214/aos/1176344136}

\bibitem[{Seidel {et~al.}(2020)Seidel, Ehrenreich, Pino, Bourrier, Lavie, Allart, Wyttenbach, \& Lovis}]{Seidel2020}
Seidel, J.~V., Ehrenreich, D., Pino, L., {et~al.} 2020, A\&A, 633, A86, \dodoi{10.1051/0004-6361/201936892}

\bibitem[{Sellke {et~al.}(2001)Sellke, Bayarri, \& Berger}]{Sellke2001}
Sellke, T., Bayarri, M.~J., \& Berger, J.~O. 2001, The American Statistician, 55, 62, \dodoi{10.1198/000313001300339950}

\bibitem[{Sing {et~al.}(2011)Sing, Pont, Aigrain, Charbonneau, D{\'e}sert, Gibson, Gilliland, Hayek, Henry, Knutson, Lecavelier Des~Etangs, Mazeh, \& Shporer}]{Sing2011}
Sing, D.~K., Pont, F., Aigrain, S., {et~al.} 2011, Monthly Notices of the Royal Astronomical Society, 416, 1443, \dodoi{10.1111/j.1365-2966.2011.19142.x}

\bibitem[{Spake {et~al.}(2021)Spake, Sing, Wakeford, Nikolov, {Mikal-Evans}, Deming, Barstow, Anderson, Carter, Gillon, Goyal, Hebrard, Hellier, Kataria, Lam, Triaud, \& Wheatley}]{Spake2021}
Spake, J.~J., Sing, D.~K., Wakeford, H.~R., {et~al.} 2021, Monthly Notices of the Royal Astronomical Society, 500, 4042, \dodoi{10.1093/mnras/staa3116}

\bibitem[{Speagle(2020)}]{Speagle2020}
Speagle, J.~S. 2020, Monthly Notices of the Royal Astronomical Society, 493, 3132, \dodoi{10.1093/mnras/staa278}

\bibitem[{Spiegelhalter {et~al.}(2014)Spiegelhalter, Best, Carlin, \& Linde}]{Spiegelhalter2014}
Spiegelhalter, D.~J., Best, N.~G., Carlin, B.~P., \& Linde, A. 2014, Journal of the Royal Statistical Society Series B: Statistical Methodology, 76, 485, \dodoi{10.1111/rssb.12062}

\bibitem[{Spiegelhalter {et~al.}(2002)Spiegelhalter, Best, Carlin, \& {van der Linde}}]{Spiegelhalter2002}
Spiegelhalter, D.~J., Best, N.~G., Carlin, B.~P., \& {van der Linde}, A. 2002, Journal of the Royal Statistical Society: Series B (Statistical Methodology), 64, 583, \dodoi{10.1111/1467-9868.00353}

\bibitem[{Swain {et~al.}(2014)Swain, Line, \& Deroo}]{Swain2014}
Swain, M.~R., Line, M.~R., \& Deroo, P. 2014, The Astrophysical Journal, 784, 133, \dodoi{10.1088/0004-637X/784/2/133}

\bibitem[{Taylor {et~al.}(2023)Taylor, Radica, Welbanks, MacDonald, Blecic, Zamyatina, Roth, Bean, Parmentier, Coulombe, Feinstein, Espinoza, Benneke, Lafreni{\`e}re, Doyon, \& Ahrer}]{Taylor2023}
Taylor, J., Radica, M., Welbanks, L., {et~al.} 2023, Monthly Notices of the Royal Astronomical Society, 524, 817, \dodoi{10.1093/mnras/stad1547}

\bibitem[{Thorngren \& Fortney(2018)}]{Thorngren2018}
Thorngren, D.~P., \& Fortney, J.~J. 2018, The Astronomical Journal, 155, 214, \dodoi{10.3847/1538-3881/aaba13}

\bibitem[{Thorngren {et~al.}(2021)Thorngren, Fortney, Lopez, Berger, \& Huber}]{Thorngren2021}
Thorngren, D.~P., Fortney, J.~J., Lopez, E.~D., Berger, T.~A., \& Huber, D. 2021, The Astrophysical Journal, 909, L16, \dodoi{10.3847/2041-8213/abe86d}

\bibitem[{Tran {et~al.}(2022)Tran, Bowler, Endl, Cochran, MacQueen, Gandolfi, Persson, Fridlund, Palle, Nowak, Deeg, Luque, Livingston, Kab{\'a}th, Skarka, {\v S}ubjak, Howell, Albrecht, Collins, Esposito, Van~Eylen, Grziwa, Goffo, Huang, Jenkins, Karjalainen, Karjalainen, Knudstrup, Korth, Lam, Latham, Levine, Osborne, Quinn, Redfield, Ricker, Seager, Serrano, Smith, Twicken, \& Winn}]{Tran2022}
Tran, Q.~H., Bowler, B.~P., Endl, M., {et~al.} 2022, AJ, 163, 225, \dodoi{10.3847/1538-3881/ac5c4f}

\bibitem[{Trotta(2008)}]{Trotta2008}
Trotta, R. 2008, Contemporary Physics, 49, 71, \dodoi{10.1080/00107510802066753}

\bibitem[{Tsai {et~al.}(2023)Tsai, Lee, Powell, Gao, Zhang, Moses, H{\'e}brard, Venot, Parmentier, Jordan, Hu, Alam, Alderson, Batalha, Bean, Benneke, Bierson, Brady, Carone, Carter, Chubb, Inglis, Leconte, Line, {L{\'o}pez-Morales}, Miguel, Molaverdikhani, Rustamkulov, Sing, Stevenson, Wakeford, Yang, Aggarwal, Baeyens, Barat, {De Val-Borro}, Daylan, Fortney, France, Goyal, Grant, Kirk, Kreidberg, Louca, Moran, Mukherjee, Nasedkin, Ohno, Rackham, Redfield, Taylor, Tremblin, Visscher, Wallack, Welbanks, Youngblood, Ahrer, Batalha, Behr, {Berta-Thompson}, Blecic, Casewell, Crossfield, Crouzet, Cubillos, Decin, D{\'e}sert, Feinstein, Gibson, Harrington, Heng, Henning, Kempton, Krick, Lagage, Lendl, Lothringer, Mansfield, Mayne, {Mikal-Evans}, Palle, Schlawin, Shorttle, Wheatley, \& Yurchenko}]{Tsai2023}
Tsai, S.-M., Lee, E. K.~H., Powell, D., {et~al.} 2023, Nature, 617, 483, \dodoi{10.1038/s41586-023-05902-2}

\bibitem[{Tsiaras {et~al.}(2019)Tsiaras, Waldmann, Tinetti, Tennyson, \& Yurchenko}]{Tsiaras2019}
Tsiaras, A., Waldmann, I.~P., Tinetti, G., Tennyson, J., \& Yurchenko, S.~N. 2019, Nature Astronomy, 3, 1086, \dodoi{10.1038/s41550-019-0878-9}

\bibitem[{Van Der~Linde(2005)}]{VanDerLinde2005}
Van Der~Linde, A. 2005, Statistica Neerlandica, 59, 45, \dodoi{10.1111/j.1467-9574.2005.00278.x}

\bibitem[{Vehtari {et~al.}(2017)Vehtari, Gelman, \& Gabry}]{Vehtari2017}
Vehtari, A., Gelman, A., \& Gabry, J. 2017, Stat Comput, 27, 1413, \dodoi{10.1007/s11222-016-9696-4}

\bibitem[{Watanabe(2010)}]{Watanabe2010}
Watanabe, S. 2010, Journal of Machine Learning Research, 11, 3571

\bibitem[{Welbanks {et~al.}(2019)Welbanks, Madhusudhan, Allard, Hubeny, Spiegelman, \& Leininger}]{Welbanks2019}
Welbanks, L., Madhusudhan, N., Allard, N.~F., {et~al.} 2019, ApJ, 887, L20, \dodoi{10.3847/2041-8213/ab5a89}

\bibitem[{Welbanks {et~al.}(2023)Welbanks, McGill, Line, \& Madhusudhan}]{Welbanks2023}
Welbanks, L., McGill, P., Line, M., \& Madhusudhan, N. 2023, The Astronomical Journal, 165, 112, \dodoi{10.3847/1538-3881/acab67}

\bibitem[{Welbanks {et~al.}(2025)Welbanks, Nixon, McGill, Tilke, Wiser, Rotman, Mukherjee, Feinstein, Line, Seager, Beatty, Seligman, Parmentier, \& Sing}]{Welbanks2025}
Welbanks, L., Nixon, M.~C., McGill, P., {et~al.} 2025, The {{Challenges}} of {{Detecting Gases}} in {{Exoplanet Atmospheres}},  arXiv, \dodoi{10.48550/arXiv.2504.21788}

\bibitem[{Xuan {et~al.}(2022)Xuan, Wang, Ruffio, Knutson, Mawet, Molli{\`e}re, Kolecki, Vigan, Mukherjee, Wallack, Wang, Baker, Bartos, Blake, Bond, Bryan, Calvin, Cetre, Chun, Delorme, Doppmann, Echeverri, Finnerty, Fitzgerald, Horstman, Inglis, Jovanovic, L{\'o}pez, Martin, Morris, Pezzato, Ragland, Ren, Ruane, Sappey, Schofield, Skemer, Venenciano, Wallace, \& Wizinowich}]{Xuan2022}
Xuan, J.~W., Wang, J., Ruffio, J.-B., {et~al.} 2022, ApJ, 937, 54, \dodoi{10.3847/1538-4357/ac8673}

\bibitem[{Yang \& Berger(1996)}]{Yang1996}
Yang, R., \& Berger, J.~O. 1996, A Catalog of Noninformative Priors ({Institute of Statistics and Decision Sciences, Duke University})

\bibitem[{Zhang {et~al.}(2021)Zhang, Snellen, Bohn, Molli{\`e}re, Ginski, Hoeijmakers, Kenworthy, Mamajek, Meshkat, Reggiani, \& Snik}]{Zhang2021a}
Zhang, Y., Snellen, I. A.~G., Bohn, A.~J., {et~al.} 2021, Nature, 595, 370, \dodoi{10.1038/s41586-021-03616-x}

\end{thebibliography}
\end{document}